\documentclass[11pt]{article}
\usepackage{aaspp4}
\newcommand{\delne}{\delta n_{\mathrm{e}}}

\newcommand{\nbar}{\overline{n}_{\mathrm{e}}}

\newcommand{\ths}{\theta_{\mathrm{s}}}

\newcommand{\cnsq}{C_n^2}

\newcommand{\smu}{\mbox{kpc~m${}^{-20/3}$}}

\newcommand{\sgra}{\mbox{Sgr~A${}^*$}}
\newcommand{\delgc}{\Delta_{\mathrm{GC}}}
\newcommand{\dgc}{D_{\mathrm{GC}}}
\newcommand{\thxgal}{\theta_{\mathrm{xgal}}}
\newcommand{\thgal}{\theta_{\mathrm{Gal}}}

\newcommand{\nexp}{\langle N\rangle}
\newcommand{\te}{T_{\mathrm{e}}}
\newcommand{\tb}{T_{\mathrm{ff}}}
\newcommand{\thgc}{\theta_{\mathrm{GC}}}
\newcommand{\cl}{\mathbf{\mathcal{L}}}
\newcommand{\ngal}{\mbox{$N_{\mathrm{Gal}}$}}
\newcommand{\nxgal}{\mbox{$N_{\mathrm{xgal}}$}}
\newcommand{\nunid}{\mbox{$N_{\mathrm{unid}}$}}

\newcommand{\va}{v_{\mathrm{A}}}

\newcommand{\hoh}{H${}_2$O}

\newcommand{\aips}{\textsc{aips}}

\lefthead{Lazio \& Cordes}
\righthead{Likelihood Analysis of Galactic Center Scattering}

\received{1997 July~23}
\paperid{36897}

\begin{document}

\title{Hyperstrong Radio-Wave Scattering in the Galactic
	Center. II. \\
	A Likelihood Analysis of Free Electrons in the
	Galactic Center}

\author{T.~Joseph~W.~Lazio\altaffilmark{1}}
\affil{Naval Research Laboratory, Code~7210, Washington, DC
	20375-5351; lazio@rsd.nrl.navy.mil}

\and

\author{James~M.~Cordes}
\affil{Department of Astronomy and National Astronomy \& Ionosphere
	Center, \\
	Cornell University, Ithaca, NY  14853-6801; \\
	cordes@spacenet.tn.cornell.edu}

\authoraddr{T. Joseph W. Lazio
            Remote Sensing Division
            NRL, Code 7210
            4555 Overlook Ave. SW
            Washington, DC  20375-5351}
\altaffiltext{1}{NRC-NRL Research Associate}

\begin{abstract}
The scattering diameters of \sgra\ and several nearby OH masers
($\approx 1\arcsec$ at 1~GHz) indicate that a region of enhanced
scattering is along the line of sight to the Galactic center.  We
combine radio-wave scattering data and free-free emission and
absorption measurements in a likelihood analysis that constrains the
following parameters of the GC scattering region: The GC-scattering
region separation, $\delgc$; the angular extent of the region,
$\psi_\ell$ and $\psi_b$; the outer scale on which density
fluctuations occur, $l_0$; and the gas temperature, $\te$.  The
maximum likelihood estimates of these parameters are $\delgc =
133_{-80}^{+200}$~pc, $0.5\arcdeg \le \psi_\ell \lesssim 1\arcdeg$,
and $(l_0/1\,\mathrm{pc})^{2/3}\te^{-1/2} = 10^{-7 \pm 0.8}$.  The
parameter $\psi_b$ was not well constrained and we adopt $\psi_b =
0\fdg5$.  The close correspondence between $\delgc$ and
$\psi_\ell\dgc$ suggests that the scattering region encloses the
\hbox{GC}.  As host media for the scattering, we consider the
photoionized surface layers of molecular clouds and the interfaces
between molecular clouds and the $10^7$~K ambient gas.  We are unable
to make an unambiguous determination, but we favor the interface model
in which the scattering medium is hot ($\te \sim 10^6$~K) and dense
($n_{\mathrm{e}} \sim 10$~cm${}^{-3}$).  The GC scattering region
produces a 1~GHz scattering diameter for an extragalactic source of
90\arcsec, if the region is a single screen, or 180\arcsec, if the
region wraps around the GC, as appears probable.  We modify the
Taylor-Cordes model for the Galactic distribution of free electrons in
order to include an explicit GC component.  We predict that pulsars
seen through this region will have a dispersion measure of
approximately $2000$~pc~cm${}^{-3}$, of which approximately
1500~pc~cm${}^{-3}$ arises from the GC component itself.  We stress
the uniqueness of the GC scattering region, probably resulting from
the high-pressure environment in the \hbox{GC}.
\end{abstract}


\setcounter{footnote}{0}

\section{Introduction}\label{sec:gc.intro}

Davies, Walsh, \& Booth~(1976) established that the observed diameter
of \sgra, the compact source in the Galactic center, scales as
$\lambda^2$, as expected if interstellar scattering from
microstructure in the electron density determines the observed
diameter.  The observed diameter of \sgra\ is now known to scale as
$\lambda^2$ from 30~cm to 3~mm (\cite{rogersetal94}) and to be
anisotropic at least over the wavelength range 21~cm to 7~mm
(\cite{bzkrml93}; \cite{krichbaumetal93}; \cite{y-zcwmr94}).  Maser
spots in OH/IR stars within 25\arcmin\ of \sgra\ also show enhanced,
anisotropic angular broadening (\cite{vfcd92}; \cite{fdcv94}).  These
observations indicate that a region of enhanced scattering with an
angular extent of at least 25\arcmin\ in radius (60~pc at 8.5~kpc) is
along the line of sight to \sgra.  At 1~GHz the level of angular
broadening produced by this scattering region is roughly 10 times
greater than that predicted by a recent model for the distribution of
free electrons in the Galaxy (\cite{tc93}, hereinafter TC93), even
though this model includes a general enhancement of scattering toward
the inner Galaxy.

These observations do not constrain the \emph{radial} location of the
scattering region for the following reason: All previous observations
have been of sources in or near the Galactic center, and for such
sources, a region of moderate scattering located far from the Galactic
center can produce angular broadening equivalent to that from a region
of intense scattering located close to the Galactic center.  Previous
estimates for the location of the scattering region have ranged
from~10~pc to~3~kpc.  Ozernoi \& Shisov~(1977) concluded that an
``unrealistic'' level of turbulence is implied unless the region is
within 10~pc of the Galactic center.  The level of turbulence they
considered unrealistic, however, namely $\sqrt{\langle
n_{\mathrm{e}}^2\rangle}/\langle n_{\mathrm{e}}\rangle \sim 1$, does
appear to occur elsewhere in the interstellar medium (\cite{s91}).
Further, van~Langevelde et al.~(1992) used the free-free absorption
toward \sgra\ to constrain the region's distance from the Galatic
center to the range 0.85--3~kpc, though suitable adjustment of free
parameters (outer scale and electron temperature) can decrease the
limit to 0.03~kpc.  We shall refer to the case in which the region is
a site of extreme scattering, $\lesssim 100$~pc from the Galactic
center and presumably caused by processes occurring there, as the GC
model.  We shall refer to the case in which the region is far from the
GC, $\gtrsim 1$~kpc and a site of enhanced but not extreme scattering,
as the random superposition (RS) model.  Although the GC model is
attractive for phenomenological reasons, other sites of enhanced
interstellar scattering are found throughout the Galaxy (e.g.,
NGC~6634, \cite{mrgb90}; Cyg~X-3, \cite{mmrj95}) and the mean free
path for encountering such a region is approximately 8~kpc
(\cite{cwfsr91}).

Identifying the location of the scattering is important in
establishing the origin of the scattering.  Associating the scattering
with a specific region may elucidate the mechanism for the generation
of the density fluctuations responsible for the scattering.  The
currently favored mechanism is that velocity or magnetic field
fluctuations---or both---generate the density fluctuations
(\cite{h84}, 1986; Montgomery, Brown, \& Matthaeus~1987; \cite{s91};
\cite{sg94}; \cite{gs95}).  Velocity or magnetic field fluctuations
are also a natural means for inducing anisotropy in the density
fluctuations and thereby in the scattering disks.  If this mechanism
is correct, the amplitude of the density fluctuations may provide a
measure of the coupling between the density and velocity or magnetic
field fluctuations or, more generally, provide information about the
small-scale velocity or magnetic field in the scattering region.
However, current observational constraints are uncertain by the ratio
of the Galactic center-scattering region distance to the Galactic
center-Sun distance.  In the RS model, the ratio is a few while in the
GC model the ratio could be as large as one hundred, so the location
of the scattering region is a key free parameter.

The location of the scattering region also has implications for pulsar
searches toward the \hbox{GC}.  Cordes \& Lazio~(1997) showed that
even if the RS model is correct, pulsars seen through the scattering
region will suffer pulse broadening of at least 5~s at 1~GHz (see also
\cite{dwb76}; \cite{os77}).  If the GC model is correct, only at
frequencies above 10~GHz will pulsations be detectable (because of the
$\nu^{-4}$ dependence of pulse broadening) and then only for pulsars
with periods longer than 100~ms.

In this paper we develop a likelihood analysis to quantify the most
probable $\delgc$ for the scattering region.  In \S\ref{sec:gc.model}
we describe our model for the distribution of free electrons in the
\hbox{GC}.  In \S\ref{sec:likefunc} we assemble measurements from the
literature relevant to radio-wave scattering and develop a likelihood
method to constrain the properties of the scattering region, and in
\S\ref{sec:gc.conclude} we discuss our results and present our
conclusions.

\section{Electron Density Model for the Galactic Center}\label{sec:gc.model}

The TC93 model synthesized scattering measurements of pulsars, masers,
and extragalactic sources and dispersion measurements of pulsars with
independently known distances.  This model for the global distribution
of free electrons in the Galaxy does not include the enhanced
scattering toward the GC, though TC93 acknowledged its existence and
recognized this deficiency in the model.  Their model underpredicts
the angular broadening of \sgra\ and nearby sources by about a factor
of 10 at 1~GHz.  In this section we augment the TC93 model by
considering the distribution of free electrons in the GC (see also
\S\ref{sec:tcmodel}).

For the GC region, the simplest model is one that can account for
the following:
\begin{enumerate}
\item Comparable, large angular broadening of \sgra\ and OH/IR masers 
to at least about 30\arcmin\ away from \sgra\ (\cite{vfcd92};
\cite{fdcv94}; \cite{y-zcwmr94});
\item Strong free-free emission and absorption within the Sgr~A
complex extending to approximately 5\arcmin\ from \sgra\
(\cite{paegvsz89}; \cite{apeg91});
\item Weak free-free emission and absorption over a wider region,
of order 10\arcmin, that includes the OH/IR masers (\cite{apeg91}); and
\item Free-free absorption of \sgra\ at frequencies below
about 0.9~GHz with an optical depth of unity in the frequency range
0.8--1~GHz (\cite{dwb76}; \cite{bdmz96}).
\end{enumerate}
To account for these observations with the simplest model, we consider
thermal free electrons distributed in two main components: (1) a
central spheroid of radius $R_{\mathrm{c}}$ centered on \sgra, and (2)
a screen at distance $\delgc$ from \sgra.  Figure~\ref{fig:geometry}
shows other geometries for the scattering region that may be
appropriate for the GC model; we can approximate the scattering region
by a screen because of the weighting factor $\delgc/\dgc$, where
$\dgc$ is the Galactic center-Sun distance (see below).  In the GC
model, the portion of the scattering region closest to the Earth will
make the largest contribution to the scattering.

We assume the electron density fluctuations have a spatial power
spectrum of the form (\cite{crcf87})
\begin{equation}
P_{\delta n_{\mathrm{e}}}(q,z) 
 = \cnsq(z) q^{-\alpha} e^{-(ql_1/2)},
\label{eqn:spectrum1}
\end{equation}
for spatial wavenumbers $q \gg q_0 = 2\pi/l_0$.  The outer and inner
scales to the density spectrum are $l_{0,1}$, respectively, and
$\cnsq(z)$ is assumed to vary slowly along the line of sight.  We
include the inner scale in our description of the spectrum because
\sgra\ is one of a small number of lines of sight for which an inner
scale may have been detected (\cite{sg90}).  The more conventional
power-law description, $P_{\delne} \propto q^{-\alpha}$ (\cite{r90};
Armstrong, Rickett, \& Spangler~1995; and references within) is
obtained for wavenumbers $q_0 \ll q \ll q_1 = 2\pi/l_1$.  We shall
also henceforth assume that the spectral index of the spectrum is
$\alpha = 11/3$, the Kolmogorov value, as suggested by a number of
observations (\cite{r90}).

A point source in the Galactic center viewed through the scattering
region, at a frequency of $\nu_{\mathrm{GHz}}$~GHz, has an apparent
diameter of (\cite{vfcd92})
\begin{equation}
\ths
 = 133\,\mathrm{mas}\,\nu_{\mathrm{GHz}}^{-2}\left(\frac{l_1}{100\,\mathrm{km}}\right)^{(4-\alpha)/2}\left[\int_0^D\cnsq(z)\left(\frac{z}{D}\right)^2\,dz\right]^{1/2}.
\label{eqn:thgal}
\end{equation}
The integral is taken from the source to the observer.  The factor
$(z/D)^2$ in equation~(\ref{eqn:thgal}) is the cause of the
aforementioned distance ambiguity (\cite{l77}; \cite{os77}; Cordes,
Weisberg, \& Boriakoff~1985; \cite{vfcd92}).  In contrast the
scattering diameter of a compact extragalactic source viewed through
this scattering region is (\cite{vfcd92})
\begin{equation}
\thxgal = \frac{\dgc}{\delgc}\thgal,
\label{eqn:xgalsize}
\end{equation}
where $\thgal$ is the characteristic diameter of a GC source---the
diameter of \sgra\ at 1~GHz is 1\farcs3---and we take the GC-Sun
distance to be $\dgc = 8.5$~kpc (at this distance $1\arcmin =
2.5$~pc).  Figure~\ref{fig:xgalsize} shows $\thxgal$ as a function of
$\delgc$.  If the RS model is correct and $\delgc \gtrsim 1$~kpc, we
expect extragalactic source diameters to be a few arcseconds; if the
GC model is correct and $\delgc \approx 100$~pc, source diameters
could exceed 1~\emph{arc\,min}.  However, few extragalactic sources
have been identified toward the \hbox{GC}.  The two closest sources
are B1739$-$298 (\cite{dkvgh83}) and GPSR~0.539$+$0.263 (Bartel~1994,
private communication), which are 48\arcmin\ and 40\arcmin\ from
\sgra, respectively.  Neither of these is within the region of
enhanced scattering defined by the OH masers.  Our observations of the
GC (Lazio \& Cordes~1997, hereinafter \cite{lc97}) revealed an
apparent deficit of sources within approximately 1\arcdeg\ of \sgra,
suggestive of extragalactic sources being broadened enough to be
resolved out by our observations.

To describe the free electron density and its fluctuations, we use the
conventional line-of-sight measures, EM, DM, and~SM (\cite{cwfsr91}).
For simplicity, we consider the statistics of the electron density to
be homogeneous in both components.  We consider the electron density
to be large in ``cloudlets'' in which the mean density is $\nbar$ and
that these cloudlets have volume filling factor $f$.  Within a
cloudlet, the fluctuations in $n_{\mathrm{e}}$ have an rms value
$\varepsilon \equiv \delta n_{\mathrm{e}} / \nbar$.  For a path length
$L$ through a medium, the three measures are
\begin{eqnarray}
DM &=& fL\nbar \nonumber \\
EM &=& fL\nbar^2 \left(1 + \varepsilon^2\right)
   = \nbar DM \left(1+\varepsilon^2\right)\\
SM = C_{\mathrm{SM}} l_0^{3-\alpha} fL\nbar^2\varepsilon^2
   &=& C_{\mathrm{SM}} l_0^{3-\alpha} \nbar \varepsilon^2 \mathrm{DM}
   = C_{\mathrm{SM}} l_0^{3-\alpha} \mathrm{EM}
            \left[ \frac{\varepsilon^2}{(1+\varepsilon^2)} \right ], \nonumber
\label{eqn:measures}
\end{eqnarray}
where $C_{\mathrm{SM}} \equiv (\alpha-3)/2(2\pi)^{4-\alpha}$ is a
constant and $\alpha$ is the spectral index of the density spectrum,
equation~(\ref{eqn:spectrum1}).

For the two components, the following general constraints are built
into our likelihood analysis:
\begin{description}
\item{\textbf{Central Component:}} From emission and absorption
measurements, we obtain the radius $R_{\mathrm{c}}$ ($\approx
5\arcmin$) and the emission measure
$\mathrm{EM}_{\mathrm{c}}(T_{\mathrm{c}})$ as a function of
temperature, $T_{\mathrm{c}}$.  From EM${}_{\mathrm{c}}$ we can
calculate $\mathrm{DM}_{\mathrm{c}}$ and $\mathrm{SM}_{\mathrm{c}}$
using the above relations.

\item{\textbf{Screen:}} Assume a thickness $\Delta d_{\mathrm{s}} \ll \delgc \ll \dgc$.
From angular diameter observations of \sgra, OH/IR masers, and the GC
transients, we determine the angular extent of the screen ($\gtrsim
15\arcmin$) and the total scattering measure as a function of
$\delgc$.  We argue that the GC scattering is dominated by the
scattering measure of the screen, $\mathrm{SM}_{\mathrm{s}}$.  From
$\mathrm{SM}_{\mathrm{s}}$, we can derive $\mathrm{DM}_{\mathrm{s}}$
and $\mathrm{EM}_{\mathrm{s}}$.
\end{description}
The overall constraints are that the total emission measure,
\begin{equation}
\mathrm{EM} = \mathrm{EM}_{\mathrm{c}} + \mathrm{EM}_{\mathrm{s}},
\label{eqn:emtot}
\end{equation}
must be dominated by the central spheroid component, since absorption
and strong emission are not seen beyond the GC spheroid, while
the weighted scattering measure
\begin{eqnarray}
\mathcal{S}(\delgc)
 & = & \int ds\, \left[ C_{n,{\mathrm{c}}}^2(s) + C_{n,{\mathrm{s}}}^2(s)\right]\left(\frac{s}{\dgc}\right)^2 \nonumber\\
 & \approx & \frac{1}{3} \mathrm{SM}_{\mathrm{c}} \left(\frac{R_{\mathrm{c}}}{\dgc}\right)^2
 + \mathrm{SM}_{\mathrm{s}} \left(\frac{\delgc}{\dgc}\right)^2
\label{eqn:smweight}
\end{eqnarray}
is dominated by the screen, since lines of sight that do not intersect
the spheroid (i.e., those toward some of the OH/IR masers) show
angular broadening at least as strong as the line of sight toward
\sgra.

The most important parameters of this model that we wish to constrain
are 
\begin{enumerate}
\item $\delgc$, the distance of scattering screen from \sgra;
\item $\psi_\ell$ and $\psi_b$, the angular extent of the scattering screen in the
$\ell$ and $b$ directions;
\item $l_0$, the outer scale of electron density fluctuations in the
screen, and 
\item $\te$, the temperature of the gas responsible for the scattering.
\end{enumerate}

\section{Likelihood Functions for Galactic Center Scattering}\label{sec:likefunc}

The measurements available for constraining scattering in the GC
consist of broadening observations of OH masers, source counts, and
free-free emission and absorption.  The joint likelihood function for
these scattering measurements is
\begin{eqnarray}
\cl &=& \cl(\theta_{\mathrm{OH}}, N, \tb, \tau_{\mathrm{ff}} | \delgc, \psi_\ell, \psi_b, l_0, \te) \nonumber \\
 &=& \cl(\tb, \tau_{\mathrm{ff}} | \theta_{\mathrm{OH}}, N; \delgc, \psi_\ell, \psi_b, l_0, \te) \cl(N | \theta_{\mathrm{OH}}; \delgc, \psi_\ell, \psi_b) \cl(\theta_{\mathrm{OH}} | \delgc, \psi_\ell, \psi_b) \nonumber \\
 &=& \cl_{\mathrm{ff}} \cl_{\mathrm{counts}} \cl_{\mathrm{broaden}}.
\label{eqn:like}
\end{eqnarray}
The various factors in this expression are 
\begin{description}
\item[$\cl_{\mathrm{broaden}}$] $= \cl(\theta_{\mathrm{OH}} |
\delgc, \psi_\ell, \psi_b)$ is the likelihood function for angular broadening
measurements of sources in and near the \hbox{GC}.  In principle we
have available measurements of OH masers (\cite{vd91}; \cite{vfcd92};
\cite{fdcv94}), \hoh\ masers (\cite{gmrs88}), the GC transients
(\cite{dwben76}; \cite{zetal92}), and extragalactic sources
(\cite{lc97}; Bartel~1996, private communication).  In practice we
shall restrict our attention to the OH masers,
$\cl_{\mathrm{broaden}} = \cl_{\mathrm{OH}}$.  Thus
far, only OH masers and the transients have been found behind the
screen.  For the OH/IR star population the three-dimensional spatial
distribution in the GC has been inferred.  We will transform this
spatial distribution into a distribution of scattering diameters.  A
similar technique cannot be utilized for the transients because the
spatial distribution of the underlying population is not
known.\footnote{
Neutron-star or black-hole binaries are probably responsible for these
radio transients.  These systems also appear as X-ray sources and the
spatial distribution of X-ray sources toward the GC has been inferred
(\cite{gks93}).  The radio transient discovered by Davies et
al.~(1976b) was indeed determined to be positionally coincident with
an X-ray source.  No such identification was made for the radio
transient discovered by Zhao et al.~(1992).}
Should this distribution become known, a similar procedure could be
used to include an additional factor in
$\cl_{\mathrm{broaden}}$.  As there are currently only two
known transients, both within 1\arcmin\ of \sgra, their contribution
to $\cl_{\mathrm{broaden}}$ would not be substantial.
Although no \hoh\ masers or extragalactic sources with measured
angular diameters have been seen through the screen, these classes of
sources can still be used to place limits on the angular extent of the
screen.

\item[$\cl_{\mathrm{counts}}$] $= \cl(N | \theta_{\mathrm{OH}};
\delgc, \psi_\ell, \psi_b)$ is the \emph{conditional} likelihood
function for counts of individual sources in our VLA fields
(\cite{lc97}).  We shall concentrate on our fields because they are
larger and deeper than those from the Columbia Plane Survey
(\cite{zhbwp90}; \cite{hzbw92}; hereinafter the CPS).

\item[$\cl_{\mathrm{ff}}$] $= \cl(\tb, \tau_{\mathrm{ff}} |
\theta_{\mathrm{OH}}, N; \delgc, \psi_\ell, \psi_b, l_0, \te)$ is the
\emph{conditional} likelihood function for free-free emission and
absorption measurements.
\end{description}

In the remaining sections, we describe each factor in $\cl$, including
the relevant data, derivation of the likelihood factor, the parameters
of the factor, and the results for that factor.  We then combine these
likelihoods to form the global likelihood.

\subsection{Angular Broadening}\label{sec:broaden}

\subsubsection{Data}

We summarize previous angular broadening measurements for sources
toward the GC in Table~\ref{tab:ang_size}, scaling the reported
angular diameter to 1~GHz assuming a $\lambda^2$ scaling, as is
appropriate for extreme scattering (\cite{vfcd92}); the distribution
of scattering diameters is shown in Fig.~\ref{fig:broaden}.  As
discussed above, we shall focus on the OH/IR stars.

Lindqvist, Habing, \& Winnberg~(1992) determined the three-dimensional
distribution of a sample of 130 OH/IR stars; about one-half of the
OH/IR stars for which scattering diameters have been determined are in
this sample.  They found the angular distribution of OH/IR stars to be
elongated in $\ell$ with an ellipticity of 0.7--0.9 and the spatial
density of OH/IR stars to be consistent with that of an isothermal
sphere, $n \propto r^{-2}$, with a centroid near the position of
\sgra\ (and IRS~16).  This distribution will enable us to derive a
likelihood function for individual OH/IR stars, a function which will
depend on $\delgc$ and the angular extent of the screen.

\subsubsection{Likelihood Factor $\cl_{\mathrm{broaden}}$ }

We model the OH/IR star distribution with the cylindrically symmetric
form
\begin{equation}
f_{r,\phi,z}(r,\phi,z) dr\,d\phi\,dz = 
      \frac{\eta}{2\pi a_r^2 a_z
       \left [ 1 + \left(r/a_r\right)^2 + \left(z/a_z\right)^2 \right ]}
       2\pi r dr\,d\phi\,dz,
\label{eqn:frpz}
\end{equation} 
where $a_r$ and $a_z$ are the scales in the distribution and $\eta$ is a
numerical factor of order unity.  Should the OH/IR stars have a 
bar-like distribution, then there would be a $\phi$ dependent
term in equation~(\ref{eqn:frpz}).

Assume that the location of the scattering region may be described by
its location $X_{\mathrm{s}}(\ell, b)$ from the GC along an
$x$-coordinate axis directed from \sgra\ to the Sun.  The $y$-axis
points toward $\ell = 270\arcdeg$ and the $z$-axis points out of the
Galactic plane.  For a radio source in the direction $\ell, b$ at
radius $r$ and angle $\phi$ from the $x$ axis, the source-screen
distance is then
\begin{equation}
\Delta = X_{\mathrm{s}}(\ell, b) - r\cos\phi. 
\label{eqn:delta}
\end{equation}
Using $y = r\sin\phi = \dgc\sin\ell$, the radius is $r =
\sqrt{(X_{\mathrm{s}} - \Delta)^2 + (\dgc\sin\ell)^2}$ and the
transformation of the distribution in $(r, \phi, z)$ to one in
Galactic coordinates and source-screen distance is
\begin{equation}
f_{\ell, b, \Delta}(\ell, b, \Delta) =
           \left(\frac{\dgc^2}{r}\right)f_{r,\phi, z}(r, \phi, z).
\label{eqn:flbdel}
\end{equation}

When calculating geometrical effects, we consider the screen to be
infinitesimally thin.  In this limit, we envision three simple shapes
for the scattering screens, cf.\ Fig.~\ref{fig:geometry}:
\begin{description}
\item[\textbf{flat screen}:] perpendicular to the line of sight to
\sgra\ and at a distance $\delgc$ from \sgra, for which
\begin{eqnarray}
\Delta & = & \delgc - r\cos\phi, \nonumber\\
X_{\mathrm{s}} & = & \delgc;
\end{eqnarray}
\item[\textbf{cylindrical screen}:] with radius $\delgc$ coaxial with the
$z$ axis:
\begin{eqnarray}
\Delta 
 & = & \left [ \delgc^2 - (\dgc\sin\ell)^2 \right ]^{1/2} - r\cos\phi, \nonumber\\
X_{\mathrm{s}} & = & \left [ \delgc^2 -  (\dgc\sin\ell)^2 \right ]^{1/2};
\end{eqnarray}
\item[\textbf{spherical screen}:] with radius $\delgc$ centered on \sgra:
\begin{eqnarray}
\Delta & = &
   \left [ \delgc^2 - (\dgc\sin\ell)^2 - (\dgc\sin b)^2 \right ]^{1/2} 
         - r\cos\phi, \nonumber\\
X_{\mathrm{s}} & = & \left [ \delgc^2 - (\dgc\sin\ell)^2 - (\dgc\sin b)^2 \right ]^{1/2}.
\end{eqnarray}
\end{description}
In these equations, we consider only the portions of the screens that
have $X_{\mathrm{s}} > 0$; i.e., for the cylindrical and spherical
screens, we ignore the portions of the screen on the far side of
\sgra.  As we noted earlier, the weighting factor $\delgc/\dgc$ means
that the portion of the screen nearest the observer will be the
dominant contribution to the scattering.  Of course, extragalactic
sources would be affected by scattering material on both the near and
far side of \sgra.

We now transform from $\Delta$ to observed scattering diameter,
$\ths$.  For sources behind the screen with separation $\Delta>0$, the
scattering diameter is simply related to that of \sgra\ ($\thgc$) as
\begin{eqnarray}
\ths(\Delta) & = & \left( \frac{\Delta}{\delgc}\right)\left(\frac{\dgc}{\dgc + \Delta - \delgc} \right)\thgc \\ \nonumber
 & \approx & \left( \frac{\Delta}{\delgc}\right) \thgc,
\label{eqn:thscreen}
\end{eqnarray}
where we consider sources near the GC such that $\vert \Delta-\delgc
\vert \ll \dgc$.  Sources in front of the screen ($\Delta < 0$) are
broadened substantially less, in accord with the predictions of the TC93
model, and we therefore assume that the contribution from
scattering material in front of the screen is negligible compared with
that from the screen.  The TC93 scattering diameter is
$\ths^{(\mathrm{TC})} (\ell, b, D)$ with
\begin{eqnarray}
D & = & \left[ \dgc^2 + (X_{\mathrm{s}}-\Delta)^2 + (\dgc\sin\ell)^2 + (\dgc\sin b)^2 - 2\dgc(X_{\mathrm{s}}-\Delta) \right]^{1/2} \nonumber\\
 & \approx & \dgc + \Delta - X_{\mathrm{s}},
\end{eqnarray}
where the approximate equality holds for sources and screen near the
\hbox{GC}.  We thus have
\begin{equation}
\ths = \cases{ 
         \displaystyle
         \thgc \left ( \frac{\Delta}{\delgc} \right ), & $\Delta>0$, behind screen; \cr
\cr
\ths^{(\mathrm{TC})} (\ell, b, D), & $\Delta<0$, not behind screen.
}
\label{eqn:thetas}
\end{equation}

Transforming from $\Delta$ to $\ths$, we find
\begin{equation}
f_{\ell, b, \ths}(\ell, b, \theta_{\mathrm{s}})
 = \cases{
           \displaystyle\left(\frac{\delgc}{\thgc}\right)
           f_{\ell, b, \Delta}(\ell, b, \Delta = \delgc\ths/\thgc),
               & $\!\Delta>0$, behind screen; \cr
\cr
\displaystyle\frac{f_{\ell, b, \Delta}(\ell, b, \Delta \approx D-\dgc + X_{\mathrm{s}})}
    {\vert\partial\ths^{(\mathrm{TC})}/\partial\Delta\vert}, & $\!\Delta<0$, not behind screen.  
}
\label{eqn:likelbth1}
\end{equation}

From equation~(\ref{eqn:flbdel}), the distribution of scattering
diameters of OH/IR stars is
\begin{equation}
f_{\ell, b, \ths}(\ell, b, \ths) 
 = \left(\frac{\delgc \dgc^2}{r\thgc}\right)f_{r,\phi, z}(r, \phi, z).
\label{eqn:likelbth2}
\end{equation}
For the specific form of equation~(\ref{eqn:frpz}) and a flat
scattering screen,
\begin{eqnarray}
& & f_{\ell, b, \ths}(\ell, b, \ths) = \nonumber \\
& & \left(\frac{\eta\delgc \dgc^2}{\thgc a_r^2 a_z}\right)
   \left\{ 1 + a_r^{-2} \left[\delgc^2\left(1-\frac{\ths}{\thgc}\right)^2
                               + \left(\ell\dgc\right)^2 \right]
                               + \left(\frac{b\dgc}{a_z}\right)^2\right\}^{-1}. \nonumber \\
& &
\label{eqn:likelbth3}
\end{eqnarray}

To form the likelihood, $\cl_{\mathrm{OH}}$, we note that
equation~(\ref{eqn:likelbth3}) applies only to sources that are behind
the screen.  Assume that the screen boundaries transverse to the line
of sight are defined by some boundary function, $\mathcal{B}(\ell,
b)$.  Sources at $\ell,b$ such that they are not seen through the
screen will have scattering angles given by the TC93 model,
$\ths^{(\mathrm{TC})} (\ell, b)$, that are much smaller than those
given by equation~(\ref{eqn:thscreen}).  In addition, there is
measurement error on the scattering diameter whose probability
distribution function is a gaussian with standard deviation
$\sigma_{\theta}$ and mean $\ths$.  This pdf should be convolved with
the pdf $f_{\ell,b,\ths}$ given above.  In all cases we will consider,
the width of $f_{\ell, b, \ths}$ as a function of $\ths$ is much wider
than the error on $\ths$, so that we will use simply
equation~(\ref{eqn:likelbth3}).

The form of equation~(\ref{eqn:likelbth3}) indicates, for $\delgc \gg
a_r$, that the likelihood of a measurement is small unless the
scattering diameter is close to that of \sgra, $\ths \approx \thgc$.
The sources measured by Frail et al.~(1994) satisfy this constraint.
The pdf also falls off in $\ell, b$ if the transverse distances from
the GC, $\ell \dgc$ and $b\dgc$, are much larger than the respective
scales, $a_r$ and $a_z$. Since highly scattered masers are seen at
least 15\arcmin\ from \sgra\ (in the longitude direction), this
suggests that $a_r$ is not significantly smaller than $\dgc \times
15\,\mathrm{arc\,min} \sim 40$~pc.  This suggests, roughly, that
$\delgc \gtrsim a_r \ge 20$~pc.  Lindqvist et al.~(1992) find $a_r \approx
50$~pc and $a_z \approx 35$~pc.

The likelihood factor for $N_{\mathrm{OH}}$ sources is then
\begin{equation}
\cl_{\mathrm{OH}}
 = \prod_{j=1}^{N_{\mathrm{OH}}} f_{\ell,b,\ths}(\ell_j,b_j,\theta_{\mathrm{s},j}).
\label{eqn:likeohir}
\end{equation}

\subsubsection{Results}\label{sec:abresults}

Figure~\ref{fig:aggregate_as} shows the angular broadening likelihood
as a function of $\delgc$ and $\psi_\ell$.  As a function of
$\psi_\ell$, this likelihood is constructed by varying $\psi_\ell$ and
using only those OH masers for which $|\ell| < \psi_\ell$.  The
likelihood function is insensitive to $\psi_b < 1\arcdeg$; allowing
$\psi_b > 1\arcdeg$ results in OH masers with large latitudes but
small longitudes to contribute to the likelihood function.  We set
$\psi_b = 0\fdg5$.

As a function of $\delgc$, $\cl_{\mathrm{broaden}} =
\cl_{\mathrm{OH}}$ shows a pronounced peak at $\delgc \approx 150$~pc.
This result is not contrary to our earlier claim in
\S\ref{sec:gc.intro} that Galactic sources cannot constrain $\delgc$.
Any \emph{given} Galactic source cannot constrain $\delgc$, because
one can adjust SM${}_{\mathrm{s}}$ and $\delgc$ in
equation~(\ref{eqn:smweight}) to produce any desired level of angular
broadening.  However, the OH masers are drawn from a Galactic
population whose spatial distribution is known.

Equation~(\ref{eqn:likelbth3}) implies that $\cl_{\mathrm{OH}}$ has a
maximum as a function of $\delgc$.  Consider a maser seen through the
scattering region, with a measured angular diameter, $\ths$, and
position $(\ell, b)$.  The likelihood for this one source is a
function of only $(\delgc/a_r)$.  For small $(\delgc/a_r)$, $f$ is
approximately linear.  However, for large $(\delgc/a_r)$, $f \propto
(\delgc/a_r)^{-2}$.  The combination leads to a peak at $(\delgc/a_r)
\sim 1$.

In combination with angular broadening measurements for other sources,
the angular broadening likelihood function can also be used to
constrain the angular extent of the scattering region.  We have
evaluated $\cl_{\mathrm{OH}}$ as a function of $\psi_\ell$ and
$\psi_b$, with $\delgc = 150$~pc.  The likelihood is dominated by the
masers OH~359.517$+$0.001, OH~359.581$-$0.240, and OH~1.369$+$1.003.
We constrain $\psi_\ell$ to $\psi_\ell \lesssim 45\arcmin$; there is
no evidence for an asymmetric distribution in $\ell$ of scattering.
The maser OH~359.517$+$0.001 is nearly 0\fdg5 from \sgra\ in
longitude, yet has a diameter 1.5 times larger than that of \sgra.
The extragalactic source B1739$-$298 ($\ell = 358.918$, $b = 0.073$)
is approximately 1\arcdeg\ from \sgra, yet has an angular diameter
smaller than that of \sgra\ (\cite{lc97}).  The screen model of
Fig.~\ref{fig:xgalsize} cannot accommodate extragalactic source
diameters smaller than that of \sgra.  Thus, the angular extent of the
scattering region to negative longitudes must be $-0.5\arcdeg >
\psi_\ell \gtrsim -1\arcdeg$.  Toward positive longitudes there are
fewer extremely heavily scattered masers.  A more severe constraint on
the angular extent is provided by \hoh\ masers and the extragalactic
source GPSR~0.539$+$0.263.  In Sgr~B ($\approx 45\arcmin$ from \sgra)
\hoh\ masers have diameters nearly five times smaller than that of
\sgra, indicating that either they are not behind the extreme
scattering region or that they are close to the screen, i.e., have a
small $\Delta$.  The upper limit on the diameter of the extragalactic
source GPSR~0.539$+$0.263 requires either $\psi_\ell \lesssim
40\arcmin$ or that $\delgc \gtrsim 3$~kpc.  Given our constraints on
$\delgc$ from above, the lack of scattering for the \hoh\ masers and
GPSR~0.539$+$0.263 is most likely due to the limited extent of the
scattering region.  However, a larger angular extent for the
scattering region is possible if the scattering screen is patchy.

The angular extent in latitude of the scattering region is less
constrained.  The maser OH~359.581$-$0.240 is 0\fdg24 from \sgra\
and has a diameter 30\% larger than \sgra; OH~1.369$+$1.003 is 1\arcdeg\
from \sgra\ with a diameter only 25\% that of \sgra.  They constrain
$\psi_b \lesssim 1\arcdeg$, but determination of angular diameters for masers (or
other sources) within the longitude range $0.25\arcdeg \lesssim |b|
\lesssim 1\arcdeg$ could provide much more stringent limits.

\subsection{Source Counts}\label{sec:counts}

Both our observations (\cite{lc97}) and those in the CPS show an
apparent deficit of sources near \sgra.  Here we quantify the
likelihood that this paucity arises from scattering so severe that
extragalactic sources have been resolved out.

\subsubsection{Data}\label{sec:countdata}

\cite{lc97} reports a VLA survey of the inner 2\arcdeg\ of the GC
designed to identify potential extragalactic sources.  We observed ten
fields at~1.28 and~1.66~GHz; the fields include one or more sources
judged likely to be extragalactic on the basis of morphology, spectral
information, or both.  We also had observations at~5~GHz, but we focus
on the 1.28 and~1.66~GHz observations because the 5~GHz observations
have smaller fields (a factor $\sim 4$ in area), so they contain fewer
sources, and the scattering diameter is no more than 10\arcsec, so
that scattering does not desensitize VLA surveys.  The observations
were conducted in spectral-line mode.  As a result a large field of
view was obtained and we were able to detect well over~100 sources.
We do not detect scatter broadening for any of these sources.

Table~\ref{tab:srccnts} shows the number of sources detected in each
field.  We report the number of Galactic, extragalactic, and
unidentified sources; anticipating the results of the next section, we
also report the number of extragalactic sources expected.

Differences in the number of sources at the two frequencies detected
in the same field arise from three effects in our survey.  A trivial,
but by far dominant, cause is the smaller field of view at 1.66~GHz.
A second, competing effect is that, in general, the rms noise in the
1.66~GHz images is slightly lower than that for the 1.28~GHz images.
Finally, the spectral index of a source could result in it being
detectable at only one frequency (a spectral index of~2 results in a
factor of~1.6 in the flux between the two frequencies).

In many cases in which a source was detected initially at only one
frequency, we have identified a possible counterpart at the other
frequency.  For the purposes of counting sources, however, we do not
consider these possible counterparts to contribute to the total number
of sources in the field, at the \emph{other} frequency.  For the sake
of specificity, we take the example of a source identified at~1.66~GHz
and a possible counterpart at~1.28~GHz.  This source would contribute
only to the 1.66~GHz counts.  Our justification for excluding the
counterpart from the source counts at~1.28~GHz is that other sources
with a similar flux were not equally likely to have been identified.
If all sources at~1.28~GHz with a flux comparable to that of the
counterpart were equally likely to have been detected, we could
describe the survey as being incomplete at a certain level at this
flux.  This is clearly not the case and we have therefore excluded
counterparts from source counts.

\subsubsection{Likelihood factor $\cl_{\mathrm{counts}}$}
\label{sec:countlike}

Within the fields observed, we detected between~2 and~20 sources per
field.  We have identified approximately 10\% of these sources as
either Galactic or extragalactic.  The remaining sources are
potentially from both Galactic and extragalactic populations.  We
compare the actual number of sources found in a field, $N$, to the
number expected, $\nexp$, using a likelihood function
\begin{equation}
\cl_{\mathrm{counts}}
 \equiv \cl(N | \theta_{\mathrm{OH}}; \delgc, \psi_\ell, \psi_b)
 = \frac{\nexp^N}{N!}\,e^{-\nexp}.
\label{eqn:countprob}
\end{equation}
The conditional nature of this likelihood function is apparent from
equation~(\ref{eqn:xgalsize}): The expected number of sources will
depend, in part, upon the expected scattering diameter for
extragalactic sources given the observed scattering diameters of
Galactic sources (i.e., OH masers and \sgra).

We shall consider a number of different determinations of
$\cl_{\mathrm{counts}}$, which differ in the way that we treat the
expected number of Galactic sources.  We discuss first the
contribution of extragalactic sources, $\langle
N_{\mathrm{xgal}}\rangle$, and then describe the various methods we
have used to estimate the Galactic source contribution.

\paragraph{Extragalactic Sources}

Within a field of radius~$\Phi$, the expected number of extragalactic
sources is
\begin{equation}
\langle N_{\mathrm{xgal}}\rangle
 = \int_0^{\Phi} d\phi\,2\pi\phi \int_{S_{\mathrm{min}}}^\infty \int_0^{\theta_{\mathrm{max}}} dS\,d\theta\,\frac{d^2n}{dS\,d\theta}.
\label{eqn:xgalnumber}
\end{equation}
Here $d^2n/dSd\theta$ is the areal density of sources on the sky per
unit flux density per unit \emph{intrinsic} diameter~$\theta$.  The limits
$S_{\mathrm{min}}$ and $\theta_{\mathrm{max}}$ on the inner integrals
result from the \emph{brightness-limited} nature of VLA surveys: A
source must be both sufficiently strong ($\ge S_{\mathrm{min}}$) and
compact ($\le \theta_{\mathrm{max}}$) to be detected.  These limits
are not constant, but, for clarity, we have suppressed the functional
dependences in equation~(\ref{eqn:xgalnumber}).

We have searched for and detected sources at considerable distances
from the field (phase) center, $\Phi \ge 20\arcmin$.  Due to the VLA's
primary beam attenuation, the minimum detectable flux density for a
point source increases with distance from the phase center,
$S_{\mathrm{min}} = S_{\mathrm{min}}(\phi)$.  At the phase centers of
the various fields, $S_{\mathrm{min}}(0)$ is 2 to~8~mJy (\cite{lc97}).

The maximum diameter for a detectable source, $\theta_{\mathrm{max}}$,
depends upon VLA configuration, flux density, and scattering diameter:
\begin{description}
\item[\textbf{VLA configuration:}] The largest angular structure which
can be detected by the VLA depends upon the length of the shortest
baselines.  For our program, the VLA was in the BnA configuration
(\cite{lc97}).  Combined with the snapshot mode of observation, the
largest detectable source diameters are approximately 60\arcsec; for
some fields near \sgra, we imposed additional $u$-$v$ constraints that
reduced this limit to~30\arcsec.

\item[\textbf{Flux density:}] Extragalactic sources show a
distribution of intrinsic angular diameter that depends on~$S$.  Even in
the absence of scattering, some sources are extended enough to escape
detection, either because the source is below our minimum detectable
brightness or because the VLA configuration is not sensitive to the
source.  The fraction of sources intrinsically large enough to avoid
detection is not negligible in the flux density range of interest.
For instance, at~10~mJy, approximately 10\% of all sources have diameters
larger than~30\arcsec\ (Windhorst, Mathis, \& Neuschaefer~1990).

\item[\textbf{Scattering diameter:}] Larger scattering diameter, $\thxgal$,
diminishes the detectability of sources, either by making sources fall
below our minimum detectable brightness or by decreasing the fraction
of sources to which the VLA configuration is sensitive.
\end{description}

We define the fraction of sources compact enough to be
detected in our program as 
\begin{equation}
f_<(\theta_{\mathrm{max}} | S; \thxgal; \mathrm{VLA})
 = \frac{\int_0^{\theta_{\mathrm{max}}} d\theta\,{d^2n}/{dS\,d\theta}}{\int_0^\infty d\theta\,{d^2n}/{dS\,d\theta}}.
\label{eqn:resolution}
\end{equation}
The quantity $\theta_{\mathrm{max}}$ is the largest \emph{intrinsic}
diameter that can be detected.  The minimum detectable brightness at
constant $S$ depends upon the maximum \emph{apparent} diameter,
$I_{\mathrm{min}} \propto S/\theta^2_{\mathrm{app,max}}$.  We take the
intrinsic and scattering diameters to add in quadrature to produce the
apparent diameter, $\theta_{\mathrm{app,max}}^2 =
\theta_{\mathrm{max}}^2 + \thxgal^2$.  We also define the areal
density of sources of all intrinsic diameters per unit flux density as
\begin{equation}
\frac{dn}{dS} = \int_0^\infty d\theta\,\frac{d^2n}{dS\,d\theta}.
\label{eqn:dnds}
\end{equation}
This quantity is reported commonly in $\log N$-$\log S$ measurements.

The expected number of extragalactic sources is then
\begin{equation}
\langle N_{\mathrm{xgal}}\rangle
 = \int_0^{\Phi} d\phi\,2\pi\phi \int_{S_{\mathrm{min}}(\phi)}^\infty dS\,f_<(\theta_{\mathrm{max}} | S; \thxgal; \mathrm{VLA})\frac{dn}{dS}.
\label{eqn:xgalnumber2}
\end{equation}
In evaluating $\langle N_{\mathrm{xgal}}\rangle$ we use the
description of the primary beam from the \aips\ task \texttt{PBCOR}
for $S_{\mathrm{min}}(\phi)$ and equation~(\ref{eqn:xgalsize}) to
calculate $\thxgal$.  We use Katgert, Oort, \& Windhorst's~(1988) fit
to 1.4~GHz source counts for $dn/dS$, and for $f_<$ we use the
intrinsic diameter distribution of Windhorst et al.~(1990).  A more
recent assessment (the FIRST survey) of both $f_<$ and $dn/dS$ shows
excellent agreement between the functional forms we have adopted and
the distributions inferred from over $10^5$ sources (\cite{wbhg97}).

We close with a caveat.  Equation~(\ref{eqn:xgalsize}) for $\thxgal$
assumes a single scattering screen.  If the scattering material wraps
around the GC, cf.\ Fig.~\ref{fig:geometry}, forming an effective
screen on both the near and far sides of the GC, the actual scattering
diameters would be at least a factor of two larger.  If the scattering
region fills the GC, the scattering could be even larger.  A larger
$\thxgal$ than we assume results in less stringent constraints on
$\delgc$.

\paragraph{Galactic Sources}\label{sec:galsources}

The second contribution to $\nexp$ is from Galactic sources.  We
consider three methods for estimating the Galactic source
contribution, $\langle\ngal\rangle$:
\begin{enumerate}
\item Assume that no Galactic sources are present;
\item Use the Galactic source distribution inferred from the CPS (\cite{hzbw92}; \cite{bwhz94}); and
\item Estimate the Galactic source distribution from the fields in our
survey.
\end{enumerate}

The Galactic radio source population is dominated by \ion{H}{2}
regions (\cite{bwmhz92}).  There are localized regions of enhanced
star formation within the GC, e.g., Sgr~B, but the star formation rate
of the inner 100~pc or so is $\lesssim 10$\% of the Galactic rate
(\cite{rg89}), and, in general, the inner 100~pc is not the site of
current, vigorous massive star formation (\cite{ms96}).  Thus,
although Method~1 will underestimate the Galactic population toward
star forming regions, on average it should not be too severe of an
underestimate of the Galactic contribution to GC fields.  The utility
of Method~1 is that it produces a minimal estimate of $\nexp$.
Method~1 therefore places an upper limit on the strength of scattering
for fields with a deficit of sources.

Extended sources (diameters $> 3\arcsec$) in the CPS are concentrated toward
both the inner Galaxy and the Galactic plane.  Becker et al.~(1992)
identify these sources with a population of extreme Population~I objects
(largely compact and ultra-compact [UC] \ion{H}{2} regions) having flux
densities $S \ge 25$~mJy.  Toward the inner Galaxy, $-20\arcdeg \le \ell
\le 40\arcdeg$, the areal density of these sources is
\begin{equation}
n_{\mathrm{Gal}}^{(2)}
 \approx 7.5\,\mathrm{deg}^{-2}\,\exp\left[-(b/0\fdg3)^2\right].
\label{eqn:galden2}
\end{equation}
Any other population of Galactic radio sources must have a scale
height larger than approximately 2\arcdeg\ (\cite{hzbw92}).  Thus, the
number of Galactic sources can be estimated from the area of the field
of view ($\sim 1$~deg${}^2$) and equation~(\ref{eqn:galden2}).  This
concentration of UC\ion{H}{2} regions to the inner Galaxy does not
contradict our assumption of Method~1 because of the different regions
involved.  The concentration toward the inner Galaxy occurs over tens
of degrees while the GC is only 1\arcdeg\ in size.

While Method~2 allows $\ngal$ to be estimated, it does require certain
caveats.  First, unlike extragalactic sources, these Galactic sources
are unlikely to be distributed randomly.  UC$\,$\ion{H}{2} regions are
often observed to be clustered within their natal molecular clouds
(\cite{c90}).  Thus, the areal density of equation~(\ref{eqn:galden2})
will likely underestimate the number of sources in fields containing
star forming regions, while overestimating the number for fields
lacking from star forming regions.  Second, the conditions for star
formation in the GC are suitably different from those in the disk
(\cite{ms96}) that it may not be valid to extend
equation~(\ref{eqn:galden2}) to fields within the \hbox{GC}.

A visual comparison shows that the fields in our survey with $|b|
\approx 1\arcdeg$ appear to contain fewer sources than those near $b =
0\arcdeg$ (\cite{lc97}).  We have assumed the high latitude fields
($|b| > 1\arcdeg$) contain only extragalactic sources while the low
latitude fields ($|b| < 1\arcdeg$) contain a mixture of extragalactic
and Galactic sources.  We have taken the Galactic sources to have a
gaussian distribution in latitude and used a maximum likelihood method
to solve for the amplitude and width of this distribution.  Using the
source counts at 1.28~GHz for those fields more than 1\arcdeg\ from
\sgra, we find
\begin{equation}
n_{\mathrm{Gal}}^{(3)}
 \approx 8.9\,\mathrm{deg}^{-2}\,\exp\left[-(b/0\fdg3)^2\right];
\label{eqn:galden3}
\end{equation}
a similar amplitude with a slightly larger width (0\fdg4) results
if we use the source counts from 1.66~GHz.

Method~3 determines $n_{\mathrm{Gal}}$ from our own data within the
region of interest, the inner few degrees, in contrast to Method~2,
which is derived from most of the inner Galaxy.  Method~3 should be
less susceptible to variations between the GC and other regions of the
Galaxy, though variations within the GC will still be important.

\subsubsection{Results}\label{sec:countresults}

We evaluate the likelihoods for the individual fields,
equation~(\ref{eqn:countprob}), assuming that the likelihoods are a
function of $\delgc$ only.  The sizes of these fields are comparable
to the known angular extent of the scattering region.  Rather than
attempt to detect changes in the number of sources as a function of
position within a field, we simply assume that a field is covered
entirely by the scattering region and compute the expected
number of sources within the field as a function of $\delgc$.

We shall obtain information about the angular extent of the scattering
region in the following manner.  We assume that the scattering
diameter of a Galactic source is that of \sgra, regardless of the
angular distance of the field from \sgra.  Since OH masers near \sgra\
have diameters comparable to that of \sgra, the scattering diameter of
\sgra\ is a reasonable description of the level of scattering, and the
expected number of sources should be roughly equal to the observed
number of sources, i.e., the likelihood should be near unity.  At
large distances from \sgra, the scattering diameter of \sgra\
(presumably) overestimates the level of scattering, and the expected
number of sources will be considerably less than the observed number,
i.e., the likelihood should be considerably less than unity.  By
determining at what angular distance the expected number of sources
becomes significantly less than the observed number, we can place
crude limits on the angular extent of the scattering region.

Figure~\ref{fig:sclike} displays the likelihoods for three fields,
359.9$+$0.2, 358.9$+$0.5, and 358.1$-$0.0.  These fields range from
15\arcmin\ to 2\arcdeg\ from \sgra\ and exemplify results for fields
at small, intermediate, and large distances from \sgra, respectively.
For the fields 358.9$+$0.5 and 358.1$-$0.0 we show the Method~3
likelihoods in which the number of Galactic sources is estimated from
our observations.  The other methods described in
\S\ref{sec:galsources} produce likelihoods with similar shapes---the
amplitudes for Method~2 are similar; for Method~1 the amplitudes are
lower.  For the field 359.9$+$0.2 we show both the Method~1 and~3
likelihoods, the Method~2 likelihood is similar to the Method~3
likelihood.

The likelihood for 358.1$-$0.0, the field farthest from \sgra, shows a
minimum at small $\delgc$, increasing with increasing $\delgc$.  If
$\delgc$ is small, this field has an excess of sources relative to
what one expects given the large scattering diameter.  For larger
$\delgc$, fewer sources are expected to be resolved out and the
likelihood increases.  The likelihood for field 358.9$+$0.5, at an
intermediate distance from \sgra, has a similar shape, though the
minimum of the likelihood is not as pronounced, indicating that the
excess is not as severe.  Finally, the likelihood for 359.9$+$0.2, the
field closest to \sgra, shows the exact opposite shape.  The
likelihood is a maximum at small $\delgc$, decreasing toward larger
$\delgc$.  This decrease reflects the increasing number of
extragalactic sources expected as $\delgc$ increases and fewer sources
are expected to be resolved out.  Thus, these likelihood functions are
consistent with the GC model, a scattering region local to the
\hbox{GC}.  The half-power point for the likelihood occurs at $\delgc
\approx 500$~pc.  The Method~1 likelihood, which compares only the
number of extragalactic and unidentified sources to the expected
number of extragalactic sources, has a maximum at intermediate
$\delgc$.  At small $\delgc$, there is an excess of sources, because
we find one source but expect none due to the extreme broadening.  At
large $\delgc$ there is a deficiency of sources as scattering is no
longer severe enough to resolve out many sources.  At intermediate
$\delgc$, $200\,\mathrm{pc} \lesssim \delgc \lesssim
700\,\mathrm{pc}$, the scattering is such that most, but not all,
extragalactic sources are expected to be resolved out.

In fact, it is likely that $\delgc < 200$~pc.  There are no
extragalactic sources in the field 359.9$+$0.2 and only one
unidentified source, 1LC~359.873$+$0.179.  We list this source as
unidentified because its morphology is suggestive of an extragalactic
source yet it does not show the level of scattering expected for an
extragalactic source seen through the entire Galactic disk
(\cite{lc97}).  If we exclude this source from the calculation of the
Method~1 likelihood, i.e., take the source to be Galactic, the paucity
of sources in this field requires $\delgc < 200$~pc.

We have performed a similar analysis for the fields from the \hbox{CPS}.  We
find the same general pattern for their fields as for ours.  Fields
far from \sgra, particularly those at latitudes $|b| > 0\fdg5$,
show likelihood minima at small $\delgc$.  Fields approximately
1\arcdeg\ from \sgra, particularly those at $b = 0\arcdeg$, show a
nearly constant likelihood.  Fields within 1\arcdeg\ of \sgra\ show
likelihood maxima at small $\delgc$.  Because of the smaller fields
($\Phi = 15\arcmin$) and generally larger $S_{\mathrm{min}}(0)$ for
the CPS fields, fewer extragalactic sources are expected and the
derived constraints on $\delgc$ are weaker than those from our fields.

Figure~\ref{fig:aggregate_sc} shows the combined source count
likelihood as a function of $\delgc$ and $\psi_\ell$, the angular
extent of the scattering screen in longitude.  We construct this
likelihood function by multiplying the likelihood functions for those
fields with $|\ell| < \psi_\ell$.  We focus on $\psi_\ell$ as a
measure of the angular extent of the screen because our fields (and
the OH/IR stars) are displaced from \sgra\ primarily in longitude.  We
fix $\psi_b = 0\fdg5$.  The likelihood function is insensitive to the
choice of $\psi_b$ for $\psi_b \lesssim 1\arcdeg$; for $\psi_b >
1\arcdeg$ the likelihood function is altered by fields with small
longitudes, but large latitudes.  The figure reflects the conclusions
we have already drawn from examining only three fields.  Over the
region $\delgc \lesssim 500$~pc and $\psi_\ell \lesssim 1\arcdeg$, the
likelihood is maximized and roughly constant.

Even though we display $\cl_{\mathrm{counts}}$ for $\psi_\ell$ as
large as 2\arcdeg, $\psi_\ell > 1\arcdeg$ is unlikely.  As discussed
in~\S\ref{sec:abresults}, the angular broadening of various
extragalactic sources and \hoh\ masers constrains the angular extent
of the region to be less than 1\arcdeg.

Two effects could make the limits on $\delgc$ less stringent.  First,
as noted in the previous section, we calculate $\thxgal$ using
equation~(\ref{eqn:xgalsize}) which assumes a single screen.  If the
scattering material wraps around the GC, so that there is a near and
far side screen, the actual scattering diameters will be at least a
factor of two larger than those calculated here.

Second, the above results assume the nominal vignetting correction
contained in the task \texttt{PBCOR}.  Zoonematkermani et al.~(1990)
suggested that \texttt{PBCOR} undercorrects flux densities at large
distances ($\Phi > 20\arcmin$) from the phase center.  If the flux
densities at large distances from the phase center are undercorrected,
the limit $S_{\mathrm{min}}(\phi)$ on the integral over $S$ in
equation~(\ref{eqn:xgalnumber}) will be too small.  Consequently, the
estimate of $\langle\nxgal\rangle$ will be too large.

The primary beam correction factor is a function of both distance from
the phase center and observing frequency.  By comparing the corrected
flux densities of sources observed in multiple fields, i.e., at
multiple distances from phase centers, Zoonematkermani et al.~(1990)
arrived at better estimates for the primary beam correction factor if
they adopted an effective frequency which was larger than their
observing frequency by about~0.2~GHz.  We repeated the above analysis
for the field 359.9$+$0.2 at~1.28~GHz, adopting an effective
frequency of 1.4~GHz.  We find a similar shape for the likelihood
function, though the upper limit on $\delgc$ is less stringent,
$\delgc \lesssim 0.7$--1~kpc.

\subsection{Free-Free Emission and Absorption Measurements}
\label{sec:free-free}

As described in \S\ref{sec:gc.model}, the density fluctuations
responsible for scattering should also contribute to the emission
measure and therefore to free-free emission and absorption.  In this
section we use angular broadening measurements to estimate
$\mathrm{EM}_{\mathrm{SM}}$ and from this, the free-free intensity,
$I_{\mathrm{SM}}$, and optical depth, $\tau_{\mathrm{SM}}$.  We then
compare these values to those derived from measurements,
$I_{\mathrm{ff}}$ and $\tau_{\mathrm{ff}}$.

\subsubsection{Data}\label{sec:gc.ffdata}

We shall focus on five masers from the Frail et al.~(1994) results in
order to predict $\mathrm{EM}_{\mathrm{SM}}$.  These masers are listed
in Table~\ref{tab:freefree}.  All have angular diameters (or geometric
means of their major and minor axes) greater than that of \sgra, all
but one show elliptical scattering disks, and all are $\ge 15\arcmin$
from \sgra.  As the free-free emission is highly concentrated toward
\sgra, these masers provide the most stringent constraints on the
amount of emission contributed by the density fluctuations responsible
for scattering.  Our use of these scattering diameters in determining
EM is the reason that the likelihood function we derive will be a
\emph{conditional} likelihood.

We estimate the free-free emission along the line of sight to these
masers from the 10~GHz survey of Handa et al.~(1987).  We apply two
corrections to the observed brightness temperature along these lines
of sight.  First, the GC contains both thermal and non-thermal
radiation.  Mezger \& Pauls~(1979) report on multi-frequency
observations, at $\nu \le 5$~GHz, from which the GC emission is
decomposed into thermal and non-thermal components.  From those
results, approximately 60\% of the 10~GHz emission is thermal.  The
second correction is that the density fluctuations responsible for
angular broadening are bounded by $\delne \le \nbar$.  Hence, at most,
only half of the thermal emission along the line of sight to the GC
can occur within the scattering region.  In summary, of the total
emission in the direction of the GC, at most 30\% of it can be
attributed to the density fluctuations in the scattering region.

We estimate the free-free absorption from the 0.327~GHz observations
of Anantharamaiah et al.~(1991).  Over the northern section of the
Arc, they estimated $\tau_{\mathrm{ff}} \approx 1$.  The fact that
they detected the Arc and Sgr~C, both at distances from \sgra\
comparable to some of the heavily scattered OH masers, indicates that
$\tau_{\mathrm{ff}}$ cannot be much larger than unity (see Fig.~2 of
\cite{fdcv94}).  At~0.16~GHz, many of the GC features cannot be
detected, indicating that $\tau_{\mathrm{ff}} > 1$ at this frequency
(\cite{y-zmsn86}).  Earlier attempts to constrain the scattering
toward the GC have focussed on the absorption toward \sgra\ only
(e.g., \cite{vfcd92}).  \sgra\ becomes obscured near 1~GHz
(\cite{dwb76}; but see also \cite{bdmz96}).  However, \sgra\ is
embedded in the Sgr~A West \ion{H}{2} region (\cite{paegvsz89};
\cite{apeg91}) and additional absorption may be contributed by this
gas.  Any additional, absorbing gas within Sgr~A West would contribute
little to the scattering of \sgra\ due to the weighting
$(R_{\mathrm{c}}/\dgc)^2$ in equation~(\ref{eqn:smweight}).  As for
the thermal emission, the density fluctuations only contribute to half
the total free-free optical depth, $\tau_{\mathrm{SM}} \le
\tau_{\mathrm{ff}}/2$.

\subsubsection{Likelihood factor $\cl_{\mathrm{ff}}$}

The EM corresponding to a source of diameter $\ths$ is (\cite{vfcd92})
\begin{equation}
\mathrm{EM}
 = 10^{2.74}\,\mathrm{pc\,cm}^{-6}\left(\frac{l_0}{1\,\mathrm{pc}}\right)^{2/3}\left(\frac{l_1}{100\,\mathrm{km}}\right)^{1/3}\left(\frac{\dgc}{\delgc}\right)^2\left(\frac{\ths}{133\,\mathrm{mas}}\right)^2.
\label{eqn:em}
\end{equation}
Although the EM does depend on the inner scale, $l_1$, this dependence
is weak compared to the other dependences in equation~(\ref{eqn:em}).
Henceforth, we will assume $l_1 = 100$~km.

Our initial attempts to form the free-free likelihood also included
the free-free absorption.  However, we found that the likelihood was
dominated by the contribution from $\tb$, reflecting the fact that we
have only upper limits on the free-free absorption.  Henceforth, we
shall restrict our attention to only the free-free emission.  The
adopted likelihood function is
\begin{equation}
\cl_{\mathrm{ff}}
 = \cl(\tb | \theta_{\mathrm{OH}}, N; \delgc, \psi_\ell, \psi_b, l_0, \te)
 = \prod_{j=1}^N \exp\left[-\left(T_{\mathrm{ff},j} - T_{\mathrm{SM},j}\right)^2/2\delta T_{\mathrm{ff},j}^2\right].
\label{eqn:pintensity}
\end{equation}
Here $T_{\mathrm{SM}}$ is the emission predicted from the angular
broadening measurements and $\tb$ is the measured quantity (with the
corrections described above).  We use the brightness temperature as a
measure of the free-free emission because that is the quantity
reported by Handa et al.~(1987).  In addition to the dependence on
$l_0$ and~$\delgc$, $T_{\mathrm{SM}}$ also depends on the electron
temperature~$\te$.

Our estimate for $\delta\tb$ reflects how accurately we can register
the maser positions in the 10~GHz survey (\cite{hsnhi87}) and the
spacing between contours.  We adopt $\delta\tb = 0.5$~K, corresponding
to $\delta\tb/\tb \approx 0.1$ to~0.5.

The predicted intensity is 
\begin{equation}
T_{\mathrm{SM}} \propto l_0^{2/3}\te^{-1/2}\delgc^{-2},
\label{eqn:predicti}
\end{equation}
where we have ignored the logarithmic dependence on temperature in the
Gaunt factor.  In contrast to the previous likelihoods in which
$\delgc$ was orthogonal to the other parameters, the free-free
likelihood cannot constrain $\delgc$ independently of $l_0$ and $\te$.

\subsubsection{Results}\label{sec:ffresults}

Figure~\ref{fig:aggregate_ff} shows $\cl_{\mathrm{ff}}$ as a function
of $\delgc$ and $l_0^{2/3}\te^{-1/2}$, with $l_0$ in parsecs and $\te$
in \hbox{K}.  As expected from equation~(\ref{eqn:predicti}), the
maximum likelihood occurs on a line of constant
$l_0^{2/3}\te^{-1/2}\delgc^{-2}$.  The corrections we have applied to
the measured brightness temperatures shift the region of maximum
likelihood to the left, i.e., to smaller $l_0^{2/3}\te^{-1/2}$ at
constant $\delgc$.

The absence of free-free absorption toward \sgra\ at frequencies above
1~GHz constrains $\delgc/\dgc$ to be $0.1 \le \delgc/\dgc \le 0.32$,
for nominal values of $l_0 = 1$~pc and $\te = 10^4$~K (\cite{vfcd92}).
The upper and lower limits scale as $l_0^{1/3}\te^{-0.675}$.  If these
nominal values are correct, the free-free emission likelihood of
Fig.~\ref{fig:aggregate_ff} indicates that $\delgc/\dgc$ would
\emph{exceed} unity.  Since the other two likelihoods are maximum for
$\delgc/\dgc < 0.1$, we anticipate that the scattering region has $l_0
< 1$~pc, $\te > 10^4$~K, or both.  We defer estimates of $l_0$ and
$\te$ until we have formed the global likelihood in the following
section.

In forming this likelihood function we have assumed that all five of
the masers in Table~\ref{tab:freefree} are affected by the scattering
screen.  The free-free likelihood does not constrain the angular
extent of the screen except that we have assumed the screen covers all
of the masers used.  The angular extent of the screen is therefore at
least 25\arcmin.

\subsection{Global Likelihood for Galactic Center 
	Scattering}\label{sec:gc.global}

Figures~\ref{fig:aggregate_as}, \ref{fig:aggregate_sc}, 
and~\ref{fig:aggregate_ff} show the individual likelihoods,
$\cl_{\mathrm{counts}}$, $\cl_{\mathrm{broaden}}$, and
$\cl_{\mathrm{ff}}$, respectively.  We now combine these to form the
global likelihood, $\cl$, equation~(\ref{eqn:like}).  This global
likelihood has four parameters, $\delgc$, $\psi_\ell$, $\psi_b$, and
$l_0^{2/3}\te^{-1/2}$.  Because $\psi_b$ is poorly constrained
(\S\ref{sec:abresults}), we shall hold it fixed at $\psi_b = 0\fdg5$.

Figure~\ref{fig:global} shows $\cl$ projected onto the
$\delgc$-$\psi_\ell$ and $\delgc$-($l_0^{2/3}\te^{-1/2}$) planes.  As
a function of $\delgc$ and $\psi_\ell$, the maximum likelihood occurs
at $\delgc = 133_{-80}^{+200}$~pc and over the range $0\fdg5 \le
\psi_\ell \le 1\fdg2$.  The ranges for these parameters enclose the
90\% confidence region and are obtained by marginalizing the global
likelihood; henceforth we adopt $\delgc = 150$~pc.  The maximum
likelihood estimate for $\delgc$ is determined primarily by the
$\cl_{\mathrm{broaden}}$ factor.  $\cl_{\mathrm{counts}}$ is constant
for $\delgc \lesssim 500$~pc while $\cl_{\mathrm{ff}}$ cannot
constrain $\delgc$.  The angular extent of the scattering region is
determined by the combination of $\cl_{\mathrm{counts}}$ and
$\cl_{\mathrm{broaden}}$.  The maximum likelihood estimate for the
angular extent is consistent with the discussion in
\S\ref{sec:abresults}, where we used individual lines of sight.

At the distance of the GC, $150\,\mathrm{pc} \approx 1\arcdeg$.  This
close correspondence between the radial and transverse sizes of the
scattering region, as given by the values of $\delgc$ and
$\psi_\ell\dgc$, indicates that the scattering region probably
encloses or fills the entire GC, rather than being a simple screen as
we have assumed, cf.\ Fig.~\ref{fig:geometry}.

With our maximum likelihood estimate of $\delgc$, we predict that the
scattering diameter for a background source seen through the GC is
90\arcsec, if the scattering is in the form of a single screen, or
180\arcsec, if the scattering region encloses the GC, as appears
probable.  This large scattering diameter makes the GC the most
extreme site of scattering known in the Galaxy; the second strongest
scattering region in the Galaxy is the \ion{H}{2} complex NGC~6634,
which produces a scattering angle of nearly 7\arcsec\ at 1~GHz
(\cite{mrgb90}).  The large scattering diameter is almost certainly a
reflection of the unique conditions in the \hbox{GC}.

As a function of $\delgc$ and $l_0^{2/3}\te^{-1/2}$, the likelihood
has a single maximum at the same $\delgc$ as above and $\log
l_0^{2/3}\te^{-1/2} = -7.0 \pm 0.8$.  If $\te = 10^4$~K, then $l_0 =
10^{-7.5}$~pc; if $l_0 = 1$~pc, then formally $\te =
10^{14}$~\hbox{K}.  The value of $l_0^{2/3}\te^{-1/2}$ in the GC is
smaller than that in the solar neighborhood, $l_0^{2/3}\te^{-1/2} \sim
10^{-2}$, i.e., $l_0 = 1$~pc and $\te = 10^4$~K (\cite{r90};
\cite{s91}; \cite{ars95}; and references within).  However, there is a
small body of evidence that indicates heavily scattered lines of sight
have $l_0 \lesssim 10^{-2}$~pc (Frail, Kulkarni, \& Vasisht~1993) and
Backer~(1978) used the scattering of \sgra\ alone to determine that a
characteristic size scale for the turbulence is approximately 5~km.
If we do not apply the corrections described in \S\ref{sec:gc.ffdata},
the maximum likelihood estimate of $l_0^{2/3}\te^{-1/2}$ increases by
about an order of magnitude.  In \S\ref{sec:gc.physical}, we consider
host media for the scattering region in light of our constraint on
$l_0^{2/3}\te^{-1/2}$.

\section{Discussion and Conclusions}\label{sec:gc.conclude}

\subsection{Comparison with Previous Analyses}

Isaacman~(1981) surveyed the central $2\arcdeg \times 4\arcdeg$ ($\ell
\times b$) of the GC in a search for planetary nebulae.  He finds an
\emph{excess} number of sources as compared to that expected from
extragalactic source counts, an excess he attributes to \ion{H}{2}
regions and planetary nebulae.  That he finds an excess at all is
notable, though, since our analysis predicts that angular diameters of
extragalactic sources seen through the GC scattering region will be at
least 1\farcm5--3\arcmin.  The resolution of his survey was $0\farcm4
\times 2\arcmin$, and he was able to detect sources with angular
scales as large as 14\arcmin.  Thus, we attribute his excess to the
fact that, where his survey overlapped the GC scattering region, it
was desensitized by the intense GC scattering to a considerably lesser
degree than our survey.

Anantharamaiah et al.~(1991) used their observations at~0.327~GHz and
a 0.408~GHz $\log N$-$\log S$ relation to conclude that the number of
observed extragalactic sources within a 4~deg${}^2$ area centered on
\sgra\ is consistent with the number expected from high-latitude
source counts.  Outside of the inner 1~deg${}^2$, our source counts
are also consistent with the expected number of extragalactic sources.
Although scattering of other sources, such as B1739$-$298 and
B1741$-$312 is heavy, the predicted diameter of these sources is less
than 10\arcsec\ at~0.327~GHz, comparable to the size of their beam, so
that these sources would not have been resolved out.

Gray et al.~(1993) surveyed the Sgr~E region ($\ell = 358.7\arcdeg, b
= 0\arcdeg$) at~0.843, 1.45, and~4.86~GHz.  At 1.4~GHz the number of
sources they find is consistent with that expected from the $\log
N$-$\log S$ distribution.  Figure~\ref{fig:sclike} shows that the
likelihood function for our field 358.9$+$0.5 does not favor a large
amount of scattering, i.e., the number of sources in this field is
consistent with that expected.  Further, two of the sources observed
in our VLBI experiment (\cite{lc97}), B1739$-$298 and
1LC~358.439$-$0.211, are in this field.  The former is heavily
scattered, though not at a level sufficient for it to be seen through
the \sgra\ scattering screen.

We conclude that our source count results are in good agreement with
previous source counts toward the \hbox{GC}.  The only exception
occurs over the 1~deg${}^2$ region centered on \sgra.  This region has
not been considered previously or has been subsumed into a much larger
area.

\subsection{Physical Conditions in the Scattering Region}\label{sec:gc.physical}

Our global likelihood, Fig.~\ref{fig:global} and \S\ref{sec:gc.global},
attained a maximum for the following parameter values: $\delgc =
150$~pc, $0.5\arcdeg \le \psi_\ell \lesssim 1\arcdeg$, and
$l_0^{2/3}\te^{-1/2} = 10^{-7}$.  In this section we consider whether
a medium exists within the GC for which such parameter values are
plausible.  There is a wealth of observational data available for the
\hbox{GC}.  We shall summarize those conclusions relevant to our study
here; interested readers are referred to a number of recent
reviews---Genzel, Hollenbach, \& Townes~(1994); Morris \&
Serabyn~(1996); and Gredel~(1996)---and references within.

Our criteria for the host medium of the density fluctuations are that
the medium must have a sufficient density and that it must be capable
of sustaining density fluctuations of the requisite magnitude.  We
establish our first criterion by estimating $\nbar$ from the
scattering diameters of GC sources using
equations~(\ref{eqn:measures}) and~(\ref{eqn:smweight}).  The
diameters of \sgra\ and the OH masers require a weighted scattering
measure of $S \approx 10^2$~\smu, equation~(\ref{eqn:smweight}).
Eliminating SM between equations~(\ref{eqn:measures})
and~(\ref{eqn:smweight}) and solving for $\nbar$ yields
\begin{equation}
\nbar
 \sim 10^3\,\mathrm{cm}^{-3}\frac{1}{\varepsilon\sqrt{f}}\left(\frac{l_0}{1\,\mathrm{pc}}\right)^{1/3}\left(\frac{\delgc}{150\,\mathrm{pc}}\right)^{-3/2}.
\label{eqn:delne}
\end{equation}

Two factors could alter this estimate by about an order of magnitude.
First, it is likely that $l_0 \ll 1$~pc, which would \emph{reduce} our
estimate of $\nbar$.  Second, as we noted in \S\ref{sec:gc.global},
the similarity between the values of $\delgc$ and $\psi_\ell\dgc$
suggests that the density fluctuations fill the region and $f \approx
1$.  However, we might also associate $l_0$ with the characteristic
size of a scattering cloudlet within the region.  If the region
contains few such cloudlets and $\delgc/l_0 \gg 1$, then $f \ll 1$,
and our estimate above would be a considerable \emph{underestimate}.
Yusef-Zadeh et al.~(1994) estimated that a typical line of sight might
intersect only 10 or so scattering cloudlets.

In any event, we conclude that the scattering medium must be dense,
$n_{\mathrm{e}} \gtrsim 10^2$~cm${}^{-3}$.  For comparison,
Spangler~(1991) concludes that $n_{\mathrm{e}} \sim 1$~cm${}^{-3}$ for
scattering regions in the Galactic disk.

Our second criterion for the host medium is that it must be able to
support density fluctuations of the required magnitude.  This
constraint has been lucidly reviewed by Spangler~(1991): The density
fluctuations are presumed to arise from plasma turbulence.  As this
turbulence dissipates, it cannot heat the host medium at a rate that
exceeds the medium's cooling capacity.  This constraint is
particularly acute in the situation we are proposing as the
dissipation mechanisms considered by Spangler~(1991) scale as
$l_0^{-a}$ with $a \approx 1$.  Since we are considering $l_0 < 1$~pc,
the heating rates could be excessive.

The dominant damping mechanisms for $l_0 < 1$~pc are linear Landau
damping, ion-neutral collisions, and a parametric decay instability.
The first two mechanisms scale as $l_0^{-2/3}$ while the latter scales
as $l_0^{-1}$.  In addition to their dependence on $l_0$, the damping
rates depend on the large scale magnetic field, $\Gamma \propto B^2$;
the Alfv\'en wave speed, $\va$; and the amplitude of the magnetic
fluctuations, $\Gamma \propto (\delta B/B)^2$ for linear Landau
damping and ion-neutral collisions while $\Gamma \propto (\delta
B/B)^3$ for the parametric decay instability.  Linear Landau damping
also depends upon the angle of propagation with respect to the
direction of $\mathbf{B}$, $\chi$, and the plasma $\beta$.  For values
of these quantities appropriate for scattering regions in the Galactic
disk, these damping mechanisms produce volumetric heating rates of
$\Gamma \sim 10^{-23.5}$--$10^{-21.5}$~erg~s${}^{-1}$~cm${}^{-3}$.  As
Spangler~(1991) discussed, there are also a number of simplifications
and additional assumptions which enter the calculation of these
heating rates.

Inferred magnetic field strengths in the GC are $B \sim 1$~mG, or
$10^3$ that of the field strength in the disk.  To estimate
$\delta B/B$, we use (Cordes, Clegg, \& Simonetti~1990)
\begin{equation}
\frac{\delne}{n_{\mathrm{e}}} \sim \left(\frac{\delta B}{B}\right)^c
\label{eqn:bB}
\end{equation}
with $c = 1$ for linear processes and $c = 2$ for non-linear processes
like the parametric decay instability.  Our first criterion for the
host medium is that $\delne/n_{\mathrm{e}} \le 1$.  For definiteness,
and to provide the largest possible value of the heating, we take
$\delta B/B \sim 1$.  Finally, although $B$ is much larger in the GC
than in the Galactic disk, $n$ is also larger.  As a result $\va$ is
larger than in the disk, but probably by no more than an order of
magnitude.  Thus, we expect $\Gamma$ in the GC to be about a factor of
$10^7$ larger that in the Galactic disk.

The heating rate from linear Landau damping in scattering regions in
the Galactic disk (\cite{s91}) assumes the density fluctuations arise
from obliquely propagating magnetosonic waves ($\chi \approx
6\arcdeg$).  More aligned propagation results in less damping.  It is
not clear if the GC environment would favor highly aligned propagation
or not.  The large values of $B$ in the GC are inferred, in part, from
the system of non-thermal filaments and threads seen throughout the
\hbox{GC}.  With only one exception, these filaments have no kinks or
bends in them, even though they are observed to be interacting with
molecular clouds having typical velocities of 10--100~km~s${}^{-1}$
(\cite{ms96}).  This rigidity could be an indication that only highly
aligned propagation is allowed.  If $\chi$ is highly concentrated near
0\arcdeg, then the heating from linear Landau damping would be
unimportant and the heating rates could be two orders of magnitude
lower than those quoted above.  Alternately, as Spangler~(1991) noted,
the distribution of $\chi$ could be isotropic, but waves with large
$\chi$ would then damp quickly and the heating rate will be unchanged
or even larger than what we assume.

The presence of small-scale ($\approx 0.1$~pc) magnetoionic cloudlets
in the GC has already been inferred to explain large changes in the
Faraday rotation measure of certain features (G~359.1$-$00.2, the
``Snake,'' \cite{gnec95}; G~359.54$+$0.18, the non-thermal filaments,
Yusef-Zadeh, Wardle, \& Parastaran~1997), though the inferred density
in these cloudlets, 0.3--10~cm${}^{-3}$, is less than our nominal
estimate.  In the Galactic disk a small body of observational evidence
suggests that the magnetoionic medium responsible for Faraday rotation
is also responsible for scattering and pulsar dispersion (\cite{sc86};
Lazio, Spangler, \& Cordes~1990; \cite{ahmsrs96}), and the same may be
true in the \hbox{GC}.

We now consider two models for the host medium.  In both models, the
scattering arises in thin layers on the surfaces of molecular clouds.
Even if the filling factor, $f$, of these layers is not large, the
\emph{covering factor}, i.e., the probability that a line of sight
through the GC will intersect one of these layers, can still be close
to unity.

\subsubsection{Photoionized Surfaces of Molecular Clouds}\label{sec:gc.skins}

Over the region $|\ell| \lesssim 1\fdg5$ and $|b| \lesssim 0\fdg5$,
$n_{\mathrm{e}} \sim 10$~cm${}^{-3}$, as determined from single-dish
recombination line and total intensity measurements (Matthews, Davies,
\& Pedlar~1973; \cite{mp79}).  Embedded within this large-scale region
are smaller regions of much higher electron densities, $n_{\mathrm{e}}
\sim 10^3$--$10^5$~cm${}^{-3}$, primarily within Sgr~A West and Sgr~B2
(e.g., \cite{mpgy-z93}).  Some of these high density regions are the
photoionized surfaces (size $\sim 10^{-4}$~pc) of molecular clouds ($n
\gtrsim 10^4$~cm${}^{-3}$) irradiated by the ambient radiation field
(effective temperature $\approx 35\,000$~K).  Yusef-Zadeh et
al.~(1994) identified these molecular skins as the source of the
scattering and associated their thicknesses with the outer scale,
$l_0$; Gray et al.~(1995) suggested that the magnetoionic medium
responsible for the Faraday rotation toward G~359.1$-$00.2 also
results from these molecular clouds.

This model suffers from at least three potential difficulties.  First,
the molecular skins are photoionized by a radiation field having a
temperature of $T_{\mathrm{eff}} \sim 10^4$~\hbox{K}.  Our constraint
on $l_0^{2/3}\te^{-1/2}$ therefore requires $l_0 \sim 10^{-7.1}$~pc.
This value is considerably smaller than that derived by Yusef-Zadeh et
al.~(1994) for the ionized molecular skins.  However, in deriving
their value for $l_0$, Yusef-Zadeh et al.~(1994) used a value of the
ionizing flux appropriate to the inner few parsecs.  The stellar
density decreases as $r^{-2}$, so outside the inner few parsecs, the
ionizing flux should be lower than that assumed by Yusef-Zadeh et
al.~(1994).  A lower ionizing flux would result in a smaller skin
depth and bring their estimate and our estimate of $l_0$ into better
agreement.

A second potential difficulty with this model is that the medium would
only barely be capable of cooling itself.  If the outer scale is $l_0
\sim 10^{-7}$~pc, then the heating rate from the damping of the plasma
turbulence is $\Gamma \sim
10^{-13}$--$10^{-12}$~erg~cm${}^{-3}$~s${}^{-1}$.  The cooling
capacity of gas near $\te \sim 10^4$~K depends sensitively upon the
fractional ionization and temperature (\cite{dm72}).  We estimate that
a density of $n_{\mathrm{e}} \gtrsim 10^5$~cm${}^{-3}$ is required for
these skins to be able to cool sufficiently in order to support the
density fluctuations.  This density is at the upper end of the range
$10^3$--$10^5$~cm${}^{-3}$ inferred for the small-scale \ion{H}{2}
regions.  In determining the cooling function of the medium, we have
used results that assume a solar abundance.  The metallicity in the GC
could be as much as twice solar, leading to an increased cooling
efficiency.

The third difficulty is that the required value for $l_0$ in this
model, $l_0 \sim 3 \times 10^{11}$~cm, is considerably smaller than
that in the Galactic disk.  In the Galactic disk, a stringent lower
limit on the outer scale is $10^{13}$~cm, and it may be as large as
$10^{18}$~cm (\cite{ars95}).  Although the physics for the generation
and maintenance of small-scale density fluctuations is not well
understood, we regard it as potentially troublesome that this model
predicts such a small $l_0$.

\subsubsection{``Warm'' Interfaces}\label{sec:gc.interface}

X-ray observations have revealed a central, diffuse X-ray source with
a (FWHM) size of $1\fdg8 \times 0\fdg9$ (\cite{ykkkt90}).  Frail et
al.~(1994) suggested that this X-ray emitting gas may be responsible
for the scattering.  Yusef-Zadeh et al.~(1997) suggested that this
X-ray emitting gas is also the magnetoionic medium responsible for the
Faraday rotation toward G~359.54$+$0.18.

The density and temperature of this region are estimated at
0.05~cm${}^{-3}$ and $10^7$--$10^8$~\hbox{K}.  This region cannot
itself be the host of the density fluctuations because of its low
density, cf.~equation~(\ref{eqn:delne}).  However, this gas appears
spatially coincident with the central zone of intense molecular
emission and presumably abuts cooler gas in the clouds.  We modify
Frail et al.'s~(1994) proposal by identifying the interfaces where the
GC molecular clouds are exposed to this ambient hot medium as the
source of the scattering.  We term these interfaces ``warm'' by
analogy with McKee \& Ostriker's~(1977) model for the \hbox{ISM}.  In
that model cold clouds immersed in a hot ($10^6$~K) medium have
$10^4$~K interfaces.  In the GC densities and temperatures are 1--2
orders of magnitude higher, but we expect that clouds will still
develop intermediate temperature interfaces.

This model suffers from two of the same difficulties as the previous
model.  The X-ray emitting gas and molecular clouds appear to be in
rough pressure equilibrium with $P \sim 5 \times 10^6$~K~cm${}^{-3}$
(\cite{bs91}).  Even though there are supersonic motions \emph{within}
the clouds, the clouds themselves ($v \sim 10$--100~km~s${}^{-1}$) are
moving subsonically with respect to the hot medium
($c_{\mathrm{sound}} \sim 1000$~km~s${}^{-1}$).  Taking pressure
balance to extend throughout the interface region, we find a density
$n_{\mathrm{e}} \sim 5$--50~cm${}^{-3}$ for $\te \sim
10^5$--$10^6$~\hbox{K}.  From our likelihood results, the temperature
within these interfaces implies an outer scale of $l_0 \sim 10^{-6.5}$--$10^{-6}$~pc, which, in turn, implies an rms density,
equation~(\ref{eqn:delne}), of $\delne \sim 10$~cm${}^{-3}$.  However,
the cooling capacity of this medium is only
$10^{-20}$~erg~cm${}^{-3}$~s${}^{-1}$.  The predicted heating rate is
$\Gamma \sim 10^{-13}$~erg~cm${}^{-3}$~s${}^{-1}$.

The outer scale in this model remains troublesomely small.  If the
size of the interface region is set by thermal conduction, the portion
of the interface with $\te < 10^6~K$ has a size $\lesssim 10^{-1}$~pc
(\cite{mc77}).  Clearly if not all of the interface contributes to the
scattering better agreement would be obtained between $l_0$ and the
interface size.  Still, the outer scale remains an order of magnitude
smaller than its lower limit in the Galactic disk.

One point in favor of this model is the distribution of the X-ray
emitting gas as compared to that of the molecular clouds.  The size of
the X-ray emitting region is similar to the extent of the scattering
region, approximately 1\arcdeg.  In contrast, the molecular cloud
distribution extends over the range $-1\arcdeg \lesssim \ell \lesssim
2\arcdeg$.  If the scattering traced massive stars within these
molecular clouds, as the photoionized molecular cloud skins model
suggests, the scattering should extend further in longitude than it
does.  In this respect, the lack of enhanced scattering for \hoh\
masers in Sgr~B is particularly problematic.

We stress the importance, and probable uniqueness, of the high density
in the \hbox{GC}.  In the Galactic disk density fluctuations cannot be
supported in media with $\te \sim 10^6$~K, a position with both
theoretical and limited observational support (\cite{s91};
\cite{pc92}).  Similarly, recent VLBI observations of 5 pulsars show
no evidence for an enhanced level of turbulence at the boundary of the
Local Bubble (\cite{cr87}; Britton, Gwinn, \& Ojeda~1996), potentially
a local analog of an interface between a hot and cooler medium.
However, the Local Bubble and ambient medium have densities a factor
of $10^2$--$10^3$ smaller than that in the \hbox{GC}.

In summary, we use our likelihood results, \S\ref{sec:gc.global}, to
constrain host media for the scattering material.  Potential media
include the photoionized skins of molecular clouds or the interface
regions between the clouds and the ambient X-ray emitting gas.  There
are difficulties with both models: Both models overpredict the outer
scale and appear to have some trouble supporting the required level of
density fluctuations.  Although we have been unable to make an
unambiguous identification of the scattering medium with either
medium, we favor the interface model, in part, because it shows a
better correspondence between the spatial distribution of scattering
and proposed host medium.

\subsection{Modification of the Taylor-Cordes Model}\label{sec:tcmodel}

The TC93 model modelled the global distribution of free electrons in the
Galaxy with four components: an extended component, an inner Galaxy
component, a component confined to the spiral arms, and a component
local to the Gum Nebula.  We now extend the TC93 model to include a
GC component\footnote{
As in TC93, the coordinate system has the $x$-axis directed parallel to
$\ell = 90\arcdeg$, the $y$-axis toward $\ell = 180\arcdeg$, the
$z$-axis toward $b = 90\arcdeg$, and $R = \sqrt{x^2 + y^2}$ is the
Galactocentric radius.} 
(cf.\ eqn.~[11] of TC93):
\begin{eqnarray}
n_{\mathrm{e}}(x,y,z) & = & n_1(R,z) + n_2(R, z) + \sum_{j=1}^4 n_{\mathrm{arm},j}(x,y,z) + n_{\mathrm{Gum}}(x,y,z) \nonumber \\
 & + & n_{\mathrm{GC}}g_{\mathrm{GC}}(R)h_{\mathrm{GC}}(z).
\end{eqnarray}
The first four components are discussed at length in TC93.  We focus on
only the last component, that toward the \hbox{GC}.

Based on the estimate in equation~(\ref{eqn:delne}) and our estimates
for $l_0$, we take $n_{\mathrm{GC}} = 10$~cm${}^{-3}$.  Our estimate
for $\delgc$ is $\delgc = 150$~pc.  Heretofore, we have been treating
the scattering region as a screen with sharp boundaries.  It is more
likely that the region has soft edges.  We therefore adopt a radial
dependence of
\begin{equation}
g_{\mathrm{GC}}(R) = \exp\left[-(R/0.150\,\mathrm{kpc})^2\right].
\end{equation}

The latitude (or $z$) dependence of the screen is less well
constrained.  For definiteness we take
\begin{equation}
h_{\mathrm{GC}}(z) = \exp\left[-(z/0.075\,\mathrm{kpc})^2\right]
\end{equation}
corresponding to $\psi_b = 0\fdg5$.  The resulting axial ratio for
the electron density distribution is 0.5; the axial ratio for the
X-ray distribution is also 0.5 and that of the molecular cloud
distribution is 0.3 (\cite{ms96}).

In the TC93 model the relationship between the free electron density
and the scattering measure produced by a line of sight of length $ds$
through those electrons is $d\mathrm{SM} \propto F
n_{\mathrm{e}}^2ds$.  The parameter~$F$ is 
\begin{equation}
F = \frac{\zeta\epsilon^2}{f}\left(\frac{l_0}{1\,\mathrm{pc}}\right)^2,
\label{eqn:F}
\end{equation}
where $\zeta$ is the normalized variance of electron density
fluctuations between cloudlets and their surroundings and $\epsilon$
and~$f$ are as in equation~(\ref{eqn:measures}).  Taking $\zeta \sim
\epsilon \sim 1$, our estimates for $l_0$ imply $F \gtrsim 10^4$.
Both of the models we have considered here have $f < 1$.  For
definiteness, we take $f \sim 0.1$, recognizing that this may be an
upper limit on~$f$.  We therefore conclude $F \gtrsim 10^5$.  For
comparison, the parameter~$F$ has a value of~0.4 in the solar
neighborhood, 6 in spiral arms, and 40 in the Galaxy's inner few
kiloparsecs.

A value of $F \sim 10^5$ produces an SM comparable to that suggested
by the scattering diameters of \sgra\ and the OH masers.  Their
scattering diameters require a line-of-sight weighted scattering
measure of $S \approx 10^2$~kpc~m${}^{-20/3}$.  Our results suggest
$\delgc \approx 150$~pc; correcting for the line-of-sight weighting,
equation~(\ref{eqn:smweight}), implies that the GC has a scattering
measure of $\mathrm{SM} \sim 10^{5.5}$~kpc~m${}^{-20/3}$.  Integrating
the TC93 expression for $d\mathrm{SM}$ through the GC, with $F \sim
10^5$, we find $\mathrm{SM} \sim 10^6$~kpc~m${}^{-20/3}$.  

Since the GC component is so localized, only for lines of sight through
the GC are the results of TC93 altered.  In this direction, however,
the TC93 model underpredicts various quantities by a large amount.  In
the TC93 model, GC pulsars have $\mathrm{DM} \approx
600$--800~pc~cm${}^{-3}$; we predict that the DM will be somewhat
larger, $\mathrm{DM} \approx 2000$~pc~cm${}^{-3}$, with approximately
1500~pc~cm${}^{-3}$ of that arising from the GC component itself.  For
comparison, the largest DM known is for PSR~B1758$-$23 with
$\mathrm{DM} = 1074$~pc~cm${}^{-3}$ (Manchester, D'Amico, \&
Tuohy~1985; \cite{klmjds93}).  Further, from Cordes \& Lazio~(1997)
the temporal broadening of pulses from pulsars seen through this
region will be $350\,\nu_{\mathrm{GHz}}^{-4}$~\emph{seconds},
requiring high-frequency ($\nu \approx 10$~GHz) periodicity searches
to detect pulsations.  Although the DM we predict for GC pulsars is
substantial, the dispersion smearing across a 1~GHz bandpass at 10~GHz
($\approx 5$~ms) is comparable to the pulse broadening, so that only
a small number of filterbank channels, e.g., 16, would be necessary to
combat the dispersion smearing.

\subsection{Conclusions}

We use a likelihood analysis to determine the following parameters of
the GC scattering region: The GC-scattering region separation,
$\delgc$; the angular extent of the region, $\psi_\ell$ and $\psi_b$;
the outer scale on which density fluctuations occur, $l_0$; and the
gas temperature, $\te$.
\begin{itemize}
\item From the literature we have assembled a list of all sources
toward the GC for which angular broadening has been measured.  A
subset of these sources is OH/IR stars, for which the spatial
distribution about the GC is known.  We construct a likelihood
function for the angular broadening of OH/IR stars, utilizing this
distribution, \S\ref{sec:broaden}.  Masers within approximately
1\arcdeg\ of \sgra\ have diameters consistent with $\delgc \approx
150$~pc (Fig.~\ref{fig:aggregate_as}).

\item The likelihood analysis of our source counts,
\S\ref{sec:counts}, indicates that a deficit of sources occurs within
approximately 1\arcdeg\ of \sgra\ and is caused by a scattering region
within 500~pc of \sgra\ (Fig.~\ref{fig:aggregate_sc}).  The resulting
scattering diameter, at least 20\arcsec\ at~1~GHz, causes
extragalactic sources to be so broad as to be resolved out by our
observations.

\item \hoh\ masers in and an extragalactic source near Sgr~B and an
extragalactic source near Sgr~C do not show the extreme scattering of
sources closer to \sgra, indicating that the scattering region does
not extend to more than 1\arcdeg\ in longitude.  The latitude extent
of the scattering region is poorly constrained, but is no more than
1\arcdeg.

\item From the literature we have estimated the free-free emission and
absorption toward five heavily scattered masers near \sgra.  The
likelihood function is dominated by the free-free emission.  The
relevant parameters, $\delgc$, $l_0$, and $\te$, are not independent
for this likelihood and only the product
$\delgc^{-2}l_0^{2/3}\te^{-1/2}$ can be constrained
(Fig.~\ref{fig:aggregate_ff}).
\end{itemize}

The global likelihood, formed by multiplying the individual
likelihoods, is shown in Fig.~\ref{fig:global}.  The maximum
likelihood estimates of the parameters are $\delgc = 150$~pc,
$0.5\arcdeg \le \psi_\ell \lesssim 1\arcdeg$, and $l_0^{2/3}\te^{-1/2}
= 10^{-7}$ with $l_0$ in pc and $\te$ in \hbox{K}.  The parameter
$\psi_b$ was not well constrained and we adopt $\psi_b = 0\fdg5$.  The
close correspondence between $\delgc$ and $\psi_\ell\dgc$ suggests
that the scattering region encloses the \hbox{GC}.

The GC scattering region produces a 1~GHz scattering diameter of
90\arcsec, if the region is a single screen, or 180\arcsec, if the
region wraps around the GC, as appears probable.  We modify the
Taylor-Cordes model for the Galactic distribution of free electrons in
order to include an explicit GC component.  We predict that pulsars
seen through this region will have a dispersion measure of
approximately $2000$~pc~cm${}^{-3}$, of which approximately
1500~pc~cm${}^{-3}$ arises from the GC component itself, and suffer
pulse broadening of $350\,\nu_{\mathrm{GHz}}^{-4}$~\emph{seconds};
pulsations will be detected only for frequencies above 10~GHz
(\cite{cl97}).

As host media for the scattering we consider the photoionized surface
layers of molecular clouds and the interfaces between molecular clouds
and the $10^7$~K ambient gas.  We identify the host medium by
requiring that it be sufficiently dense to support density
fluctuations of the required magnitude.  We are unable to make an
unambiguous determination, but we favor the interface model which
predicts that the scattering medium is hot ($\te \sim 10^6$~K) and
dense ($n_{\mathrm{e}} \sim 10$~cm${}^{-3}$).  The X-ray interface
model also shows better spatial agreement, when compared to the
photoionized skin model, with the region over which the scattering is
observed.  This model is summarized graphically in
Fig.~\ref{fig:gc.summary}.

The GC scattering region is likely to be unique in the Galaxy, probably
because it is a high-pressure environment and can sustain densities and
temperatures much higher than in the Galactic disk.

\acknowledgements
We thank Z.~Arzoumanian, D.~Frail, M.~Goss, H. van~Langevelde,
F.~Yusef-Zadeh, D.~Chernoff, P.~Goldsmith, and T.~Herter for helpful
and illuminating discussions.  We thank N.~Bartel for contributing
additional angular broadening measurements.  We thank the referee for
a suggestion that helped improve the discussion of the likelihoods.
This research made use of the Simbad database, operated at the CDS,
Strasbourg, France.  This research was supported by the NSF under
grant AST~92-18075 and AST~95-28394.  The NAIC is operated by Cornell
University under a cooperative agreement with the \hbox{NSF}.
TJWL holds a National Research Council-NRL Research Associateship.
Basic research in astronomy at the Naval Research Laboratory is
supported by the Office of Naval Research.

\clearpage

\begin{figure}
\caption[]{Possible scattering geometries toward the Galactic center.
The dark circle represents a GC source such as \sgra, and the lightly
colored regions indicate possible configurations for the ionized gas
responsible for the radio-wave scattering.  In general, we shall
approximate the scattering region by a single screen.}
\label{fig:geometry}
\end{figure}

\begin{figure}
\caption[Extragalactic Source Diameters]
{The diameter of an extragalactic source at~1.4~GHz seen through the
scattering region in front of \sgra\ as a function of the distance of
the scattering region from the Galactic center, $\delgc$.  The dotted
line indicates an extreme lower limit on $\delgc$ as derived from the
lack of free-free absorption toward \sgra\ at centimeter wavelengths.
At~1.4~GHz, the scattering diameter of \sgra\ is 0\farcs7.  The
scattering diameter scales as $\ths \propto \nu^{-2}$.}
\label{fig:xgalsize}
\end{figure}

\begin{figure}
\caption[]{Angular broadening measurements toward the GC compiled from
\cite{lc97} and the literature.  Crosses show the relative diameter of
the major and minor axes of sources with measured angular diameters.
The large star shows the relative size of the upper limit on the
scattering diameter of the extragalactic source GPSR~0.539$+$0.263.
Source diameters are scaled up by a factor of 500.  The contours are
from the 5~GHz survey with the NRAO 91~m telescope (Condon, Broderick,
\& Seielstad~1991).}
\label{fig:broaden}
\end{figure}

\begin{figure}
\caption[The Global Likelihood Function for Angular Broadening Measurements]
{$\cl_{\mathrm{broaden}}$, the likelihood function from
angular broadening measurements on OH masers  as a function of
$\delgc$ and $\psi_\ell$.  Contours show the 67\% (\emph{heavy}),
90\%, and 99\% confidence regions.}
\label{fig:aggregate_as}
\end{figure}

\begin{figure}
\caption[Likelihood Function for Source Counts in Individual Fields]
{$\cl_{\mathrm{counts}}$, the likelihood function as a function of
$\delgc$ for the fields 359.9$+$0.2 (\textit{solid}, 15\arcmin\ from
\sgra), 358.9$+$0.5 (\textit{dashed}, 77\arcmin\ from \sgra), and
358.1$-$0.0 (\textit{dot-dash}, 113\arcmin\ from \sgra).  These
likelihood functions were all computed using Method~3, which uses our
observations to estimate the number of Galactic sources in each field.
Also shown is the likelihood function for the field 359.9$+$0.2
(\textit{dotted}) in which only the number of extragalactic and
unidentified sources is compared to the number of extragalactic
sources expected.  The discontinuity occurs where the number of
detectable extragalactic sources becomes non-zero.}
\label{fig:sclike}
\end{figure}

\begin{figure}
\caption[The Global Likelihood Function for Source Counts]
{$\cl_{\mathrm{counts}}$, the likelihood function from source
counts as a function of $\delgc$ and $\psi_\ell$.  Contours show the 67\%
(\emph{heavy}), 90\%, and 99\% confidence regions.}
\label{fig:aggregate_sc}
\end{figure}

\begin{figure}
\caption[]{$\cl_{\mathrm{ff}}$, the likelihood function from
free-free emission and absorption measurements as a function of
$\delgc$ and $l_0^{2/3}\te^{-1/2}$.  This likelihood is dominated by
the free-free emission measurements.  Contours show the 67\%
(\emph{heavy}), 90\%, and 99\% confidence regions.}
\label{fig:aggregate_ff}
\end{figure}

\begin{figure}
\caption[]{The global likelihood for GC scattering.  Contours show the
67\%, 90\%, and 99\% confidence regions.  We assume $\psi_b = 0\fdg5$.
(\textit{a}) ${\cl}$ as a function of $\delgc$ and $\psi_\ell$.  The
sharp edges in the contours reflect the underlying granularity of the
data.
(\textit{b}) ${\cl}$ as a function of $\delgc$ and
$l_0^{2/3}\te^{-1/2}$, with $l_0$ in pc and $\te$ in \hbox{K}.  The
maximum likelihood occurs at $\log l_0^{2/3}\te^{-1/2} = -7.0$ and
$\log\delgc = -0.88$.}
\label{fig:global}
\end{figure}

\begin{figure}
\caption[Graphical Summary of the Galactic Center Scattering Region]
{A graphical summary of the Galactic center scattering region (not to
scale).  The lightly colored region is the FWHM extent of the X-ray
emitting ($10^7$~K) gas.  The heavily colored regions are the
molecular clouds ($10^2$~K), which are surrounded by the scattering
gas ($10^6$~K).  The central object is \sgra\ and the $+$ signs represent
some of the OH/IR stars.
}
\label{fig:gc.summary}
\end{figure}

\clearpage

\begin{deluxetable}{lccc}
\tablecaption{Angular Diameter Measurements from the
	Literature\label{tab:ang_size}}
\tablehead{\colhead{Name} & \colhead{$\theta_{\mathrm{obs}}$\tablenotemark{a}}
	& \colhead{$\Delta\Theta_{\mathrm{Sgr~A^*}}$} & \colhead{Ref.} \\
	& \colhead{(milliarcsecond)} & \colhead{(\arcmin)} }

\startdata

OH~353.298$-$1.537    	   & \phn935 & 408.6    & 5 \nl
OH~355.641$-$1.742    	   & \phn855 & 277.5    & 5 \nl
OH~355.897$-$1.754    	   & \phn904 & 263.5    & 5 \nl
OH~357.849$+$9.744    	   & \phn408 & 600.6    & 5 \nl
OH~359.140$+$1.137    	   & \phn699 & \phn85.8 & 5 \nl

\nl

OH~359.517$+$0.001    	   & 2130    & \phn25.8 & 6 \nl
OH~359.564$+$1.287    	   & \phn232 & \phn83.2 & 5 \nl     
OH~359.581$-$0.240    	   & 1680    & \phn24.7 & 6 \nl
OH~359.762$+$0.120\tablenotemark{b}  
			   & 1710    & \phn14.8 & 6 \nl
OH~359.880$-$0.087    	   & 2760    & \phn\phn4.6 & 6 \nl

\nl

OH~359.938$-$0.077    	   & 1480    & \phn\phn\phn1.9 & 3 \nl
OH~359.954$-$0.041    	   & 1540    & \phn\phn\phn0.7 & 3 \nl
OH~359.986$-$0.061    	   & 2460    & \phn\phn2.6 & 6 \nl
OH~{\phd\phd}0.125$+$5.111 & \phn100 & 309.6    & 5 \nl
OH~{\phd\phd}0.190$+$0.036 & \phn268 & \phn\phn15.6 & 3 \nl

\tableline
\tablebreak

OH~{\phd\phd}0.319$-$0.040 & 3120    & \phn22.5 & 6 \nl
OH~{\phd\phd}0.334$-$0.181 & 3160    & \phn24.7 & 6 \nl
OH~{\phd\phd}0.892$+$1.342\tablenotemark{c}
	                   & \phn163 & 100.9    & 5 \nl
OH~{\phd\phd}1.369$+$1.003 & \phn291 & 106.2    & 5 \nl
OH~{\phd\phd}1.212$+$1.258 & \phn670 & 109.1    & 5 \nl

\nl

OH~{\phd\phd}3.234$-$2.404 & \phn366 & 242.8    & 5 \nl
OH~{\phd\phd}5.026$+$1.491 & \phn194 & 318.5    & 5 \nl

\tableline
\tablebreak

\sgra             	   & 1300    & \phn\phn\phn0.0 & 7 \nl
GCT                   	   & 1310    & \phn\phn\phn0.6 & 4 \nl %
A~1742$-$28           	   & 1380    & \phn\phn\phn0.9 & 1 \nl %
Sgr~B2(N)                  & \phn276 & \phn\phn45      & 5 \nl %
GPSR~0.539$+$0.263         & $<$3900 & \phn\phn40 & 8 \nl

\enddata

\tablenotetext{a}{All diameters are scaled to 1~GHz assuming a
$\lambda^2$ scaling.  For sources with anisotropic scattering disks, we
quote the geometric mean of the semi-major and -minor axes.}
\tablenotetext{b}{Frail et al.'s~(1994) diameter, also observed by
van~Langevelde \& Diamond~(1991) and van~Langevelde et al.~(1992)}
\tablenotetext{c}{van~Langevelde et al.'s~(1992) diameter; Frail et
al.~(1994) cite difficulties with their fitting algorithm.}
\tablerefs{(1)~Davies et al.~(1976a); (2)~Gwinn et al.~(1988);
(3)~van~Langevelde \& Diamond (1991); (4)~Zhao et al.~(1991);
(5)~van~Langevelde et al.~(1992); (6) Frail et al.~(1994);
(7)~Yusef-Zadeh et al.~(1994); (8) Bartel~(1996, private
communication)}
\end{deluxetable}

\begin{deluxetable}{lcccccccc}
\tablecaption{Source Counts and Expected Source
	Numbers\label{tab:srccnts}}

\tablehead{ & \multicolumn{4}{c}{1.28~GHz}
	& \multicolumn{4}{c}{1.66~GHz} \\
	\cline{2-5} \cline{6-9}
	\colhead{Field}
	& \colhead{\ngal} & \colhead{\nxgal} & \colhead{\nunid} & \colhead{$\langle\nxgal\rangle$}
	& \colhead{\ngal} & \colhead{\nxgal} & \colhead{\nunid} & \colhead{$\langle\nxgal\rangle$}}

\startdata	

357.9$-$1.0       & \phn0 & 2 & \phn9 & 5.1 & \phn0 & 1 & \phn9 & \phn5.2 \nl
358.1$-$0.0  	  & \phn3 & 0 &    13 & 5.8 & \phn1 & 0 &    16 & \phn6.9 \nl
358.7$-$0.0       & \phn2 & 1 &    19 & 3.4 & \phn2 & 1 &    17 & \phn4.0 \nl
358.9$+$0.5  	  & \phn0 & 1 &    12 & 5.4 & \phn0 & 1 &    10 & \phn6.6 \nl
359.9$+$0.2       & \phn1 & 0 & \phn1 & 7.5 &    20 & 0 & \phn1 & \phn1.1 \nl

\nl
				  
\phn\phn0.0$+$0.0 & 20    & 0 & \phn1 & 0.4 & 28    & 0 & \phn1 & \phn3.2 \nl
\phn\phn0.2$-$0.7 & \phn0 & 0 &    14 & 2.7 & \phn0 & 0 &    13 & \phn6.7 \nl
\phn\phn0.5$+$0.2\tablenotemark{a} 
		  & \phn0 & 0 &    16 & 7.8 & \phn0 & 0 &    22 & \phn8.5 \nl
\phn\phn1.0$+$1.6 & \phn0 & 0 &    15 & 9.7 & \phn0 & 0 &    14 &    11.0 \nl
\phn\phn1.2$-$0.0 & \phn1 & 0 & \phn3 & 2.2 & \phn1 & 0 & \phn5 & \phn3.9 \nl

\enddata
\tablenotetext{a}{Extent of field limited to avoid Sgr~B.}
\end{deluxetable}

\begin{deluxetable}{lccc}
\tablecaption{Free-Free Emission Toward	Heavily Scattered OH
	Masers\label{tab:freefree}} 
\tablehead{\colhead{Name} 
	& \colhead{$\Delta\Theta_{\mathrm{Sgr~A^*}}$}
	& \colhead{$\theta_{\mathrm{1.6~GHz}}$}	& \colhead{$\tb$} \\
	& \colhead{(\arcmin)}
	& \colhead{(\arcsec)} & \colhead{(K)} \\
	\colhead{(1)}
	& \colhead{(2)}
	& \colhead{(3)}	& \colhead{(4)}}

\startdata

OH~359.517$+$0.001 	 & 25.8 & 0.82 &  0.19 \nl
OH~359.581$-$0.240 	 & 24.7 & 0.64 &  0.11 \nl
OH~359.762$+$0.120 	 & 14.8 & 0.53 &  0.14 \nl
OH~\phn\phn0.319$-$0.040 & 22.5 & 1.20 &  0.48 \nl
OH~\phn\phn0.334$-$0.181 & 24.7 & 1.22 &  0.27 \nl

\enddata

\tablecomments{(2) Angular separation between maser and \sgra; (3)
Angular diameter at 1.612~GHz, geometric mean of major and minor axis
for those masers with anisotropic scattering disks, the
scattering disk of \sgra\ at~1.612~GHz is 0\farcs5; (4)
Brightness temperature at 10~GHz, from Handa et al.~(1987), corrected
for non-thermal emission.}

\end{deluxetable}

\end{document}